\documentclass[fleqn,usenatbib]{mnras}

\usepackage{newtxtext,newtxmath}

\usepackage[T1]{fontenc}

\DeclareRobustCommand{\VAN}[3]{#2}
\let\VANthebibliography\thebibliography
\def\thebibliography{\DeclareRobustCommand{\VAN}[3]{##3}\VANthebibliography}

\usepackage{graphicx}	
\usepackage{amsmath}	
\usepackage{amssymb}	

\title[Small Lensed $\MakeLowercase{z}\geq5.5$ Galaxies]{RELICS: Small Lensed $\MakeLowercase{z}\geq5.5$ Galaxies Selected as Potential Lyman Continuum Leakers}

\author[C. Neufeld et al.]{
Chloe Neufeld,$^{1,2}$
Victoria Strait,$^{3,1}$
Maru{\v s}a Brada{\v c},$^{4,1}$
Brian C. Lemaux,$^{5,1}$ 
Dan Coe,$^6$
Lilan Yang, $^{7,8}$ 
\newauthor
Tommaso Treu, $^8$
Adi Zitrin, $^9$
Mario Nonino, $^{10}$
Larry Bradley, $^{6}$
Keren Sharon $^{11}$
\\
$^{1}$Physics and Astronomy Department, University of California, Davis, CA 95616, USA\\
$^2$Astronomy Department, Yale University, New Haven, CT 06511,
USA\\
$^3$Cosmic Dawn Center (DAWN), Niels Bohr Institute, University of Copenhagen, Denmark \\
$^4$Department of Mathematics and Physics, University of Ljubljana, Jadranska ulica 19, SI-1000 Ljubljana, Slovenia \\
$^5$Gemini Observatory, NSF’s NOIRLab, 670 N. A’ohoku Place, Hilo, Hawai’i, 96720, USA \\
$^{6}$Space Telescope Science Institute, Baltimore, MD 21218, USA\\
$^{7}$School of Physics and Technology, Wuhan University, Wuhan 430072, China\\
$^8$Department of Physics and Astronomy, University of California, Los Angeles, CA 90095-1547, USA\\
$^9$Department of Physics, Ben-Gurion University, Be’er-Sheva 84105, Israel\\
$^{10}$INAF – Osservatorio Astronomico di Trieste, via G. B. Tiepolo 11, I-34131 Trieste, Italy \\
$^{11}$Department of Astronomy, University of Michigan, 1085 South University Ave, Ann Arbor, MI 48109, USA
}

\date{Accepted XXX. Received YYY; in original form ZZZ}

\pubyear{2021}

\begin{document}
\label{firstpage}
\pagerange{\pageref{firstpage}--\pageref{lastpage}}
\maketitle

\begin{abstract}
We present size measurements of 78 high-redshift ($z\geq 5.5$) galaxy candidates from the Reionisation Lensing Cluster Survey (RELICS). These distant galaxies are well-resolved due to the gravitational lensing power of foreground galaxy clusters, imaged by the \textit{Hubble Space Telescope} (\textit{HST}) and the \textit{Spitzer Space Telescope}. We compute sizes using the forward-modeling code \textsc{Lenstruction} and account for magnification using public lens models. The resulting size-magnitude measurements confirm the existence of many small galaxies {with effective radii $R_{\rm{eff}}<200$ pc} in the early universe, in agreement with previous studies. In addition, we highlight compact and highly star-forming sources {with star formation rate surface densities $\Sigma_\text{SFR}>10M_\odot\text{yr}^{-1}\text{kpc}^{-2}$} as possible Lyman continuum leaking candidates that could be major contributors to the process of reionisation. Future spectroscopic follow-up of these compact galaxies (e.g., with the \textit{James Webb Space Telescope}) will further clarify their role in reionisation and the physics of early star formation. 
\end{abstract}




\section{Introduction}

The study of high-redshift galaxies is a key aspect of constructing a complete picture of galaxy formation and evolution. Their properties, such as size and magnitude, can lend insight into conditions in the galaxies themselves and between galaxies in the intergalactic medium (IGM) during the Epoch of Reionisation, which marks an important phase change in the universe when neutral gas was ionised by the first sources of light {\citep{dayal2019,robertson2021}}. By studying these early galaxies' properties, we can address questions involving which specific objects were responsible for reionisation and how the process occurred over time.

Knowledge of the size evolution of galaxies allows deeper understanding of galaxy evolution and formation. Previous works have studied the relation between size and luminosity (\citealp{Huang2013,Ono2013,Holwerda2015,Shibuya2015,Shibuya2019}) as well as size and stellar mass (\citealp{Franx2008,vanderWel2014,Morishita2014,Mowla2019}). These relations involve implications for the UV Luminosity Functions (LFs), especially the faint end slope {(\citealp{Grazian2011,Bouwens2017,Kawamata2018,Bouwens2021})}, as extremely small sizes at high redshifts imply a steep size-luminosity relation and a shallow faint-end slope for the LF (\citealp{Kawamata2018}). {Lensed high-redshift galaxies have also aided in constraining the faint end of the UV luminosity function (e.g., \citealp{Atek2018,Ishigaki2018,Yue2018,Bhatawdek2019}).} The faint end slope of the LF and determining the relation of galaxy sizes to their ionising photon production in the early universe are important aspects for understanding the sources of reionisation (\citealp{Grazian2012}). {Previous studies (e.g., \citealp{Atek2015,Kawamata2015,Livermore2017}) measure the faint end of the LF and find that, under certain assumptions, there are likely enough faint galaxies to} balance the ionising photon budgets, {while others (e.g., \citealp{Naidu2020}, \citealp{Mathee2021}) argue that perhaps only bright galaxies are sufficient for reionization.} In addition, \cite{Huang2017} {and \cite{Shibuya2015}} find that the sizes of galaxies are proportional to the sizes of their dark matter halos, and it has been shown that size decreases at higher redshifts for fixed luminosity or stellar mass (\citealp{Ono2013,vanderWel2014, Morishita2014,Shibuya2015,Shibuya2019,Mowla2019}), although see \cite{Ribeiro2016} for an alternate view. 

Sizes of galaxies have also been used to select for Lyman continuum leaking galaxies, likely major contributors to the process of reionisation. As found by \cite{Marchi2018}, galaxies at redshift $3.5\leq z\leq 4.3$ that are compact, with $R_{\rm{eff}} \leq 0.3$ kpc, tend to have higher inferred Lyman continuum flux (i.e., flux at $\lambda < 912$\AA\ equivalent width) than more extended sources as well as strong Lyman-$\alpha$ emission. Others have also found that Lyman continuum leakers are compact and highly star forming {(e.g., \citealp{cardamone2009,Izotov2016b,Verhamme2017,Naidu2021,Kim2021,Flury2022}, although see \citealp{Saxena2021})}, and thus measuring the sizes of galaxies in the Epoch of Reionisation is key to identifying likely leakers. 

In this work, we study the sizes of a sample of 78 star-forming galaxies at $z\geq 5.5$ using the Reionisation Lensing Cluster Survey (RELICS, PI Coe) and companion survey \textit{Spitzer}-RELICS (S-RELICS, PI Brada{\v c}), which provide imaging data of lensed high-redshift galaxy candidates. The data allows us to measure high-quality photometric redshifts estimated from a combination of \textit{Hubble Space Telescope} (\textit{HST}) and \textit{Spitzer}/IRAC data (\citealp{Coe2019,Salmon2020,Strait2020,Strait2021}). The RELICS survey is focused on observing apparently bright, lensed, high-redshift galaxies, which are optimal for detailed follow-up study. Without the help of gravitational lensing, these galaxies would otherwise be difficult to study extensively or even observe. While other \textit{HST} surveys of clusters such as the Cluster Lensing And Supernova survey (CLASH; \citealp{Postman2012}) and the Frontier Fields (\citealp{Lotz2017}) obtain deeper imaging, the shallower but wider covering area of RELICS yields brighter high-redshift candidate galaxies (\citealp{Coe2019}). RELICS also provides a lower sample variance than other surveys because it has 41 independent lines of sight, thus decreasing uncertainty due to cosmic variance. Because RELICS probes about an order of magnitude in brightness ($L^*_{UV}$ at these redshifts is roughly $M_{AB,UV}=-21$), our sample is comprised of a combination of sources at or below the characteristic luminosity $L^*_{UV}$, as well as a few galaxies brighter than $L^*_{UV}$ (see \citealp{Strait2021}). Analysis of this sample will thus complement previous discussions on the contributions of different galaxy populations to reionisation (e.g., \citealp{Finkelstein2019,Shen2020,Naidu2021}) by helping to illuminate the contributions of "medium bright" galaxies. 

Measuring the sizes of high redshift galaxies has proven to be a difficult task. Most lie at the detection limit of \textit{HST} in observations of a typical unlensed field, and thus galaxies magnified by gravitational lensing are especially important to analyse these populations, due to their increased apparent brightness and size resolution (e.g. \citealp{Kawamata2015,Kawamata2018,Atek2018,Yang2020a,Bouwens2021}). Faint {galaxies} at high redshifts have been found to have sizes (measured using effective radius $R_{\rm{eff}}$) on the order of a few hundred parsecs (\citealp{Ono2013,Vanzella2020,Bouwens2021b}), and galaxies at very high redshifts (e.g., \citealp{Coe2013,Zitrin2014}) have sizes of $\lesssim 0.03$" ($\lesssim 0.1$ kpc) at $z \sim 10$. These small sources therefore could barely be resolved even at the resolving power of \textit{HST} without gravitational lensing.

Lensed galaxies must be reconstructed to obtain their intrinsic morphology and to measure their size. To perform this reconstruction we utilize the \textsc{Python} package \textsc{\textsc{Lenstruction}}, developed by \cite{Yang2020} and used by \cite{Yang2020a} to study $z=1-3$ galaxies lensed by the \textit{HFF}. This package is built on the software \textsc{Lenstronomy} (\citealp{Birrer2015,Birrer2018}) and adopts a forward modeling technique which maps sources to the image plane, and the modeled image is compared to the observation to constrain the source light profile.

We outline the paper as follows. In Section \ref{data}, we report the specifics of our data and sample selection. We then describe our method for measuring the sizes of galaxies, including the modelling process and uncertainties, in Section \ref{sizes}. Our results are presented in Section \ref{results}, with comparisons to previous works and the selection of possible Lyman continuum leakers. In Section \ref{concls}, we provide a summary and conclusion.

We assume a $\Lambda$CDM cosmology with parameters $H_0=70$ km s$^{-1}$ Mpc $^{-1}$, $\Omega_m=0.3$, and $\Omega_\Lambda = 0.7$. All magnitudes are in the AB system \citep{Oke1974}, and all sizes are measured in proper distances. All equivalent width measurements are given in the rest frame.

\section{Data}\label{data}

\subsection{Observations and Photometry}\label{obsphot}
Throughout this work we use data from the 188-orbit Hubble Treasury Program Reionisation Lensing Cluster Survey (RELICS Cycle 23; GO 14096; PI Coe, \citealp{Coe2019}). The initial high-redshift sample was identified using \textit{HST} imaging by \cite{Salmon2020}. Here, we focus on the high-redshift candidates with reliable \textit{Spitzer} fluxes, as defined by \cite{Strait2020}. 

All targets in this work were observed to two orbit depth in WFC3/IR imaging (F105W, F125W, F140W, and F160W) and to three orbit depth in ACS (F435W, F606W, F814W), using a combination of RELICS and archival data. For initial object selection, we use the catalogues based on a detection image comprised of the weighted stack of all WFC3/IR imaging with $0.06 \arcsec$ pix$^{-1}$ resolution, optimized for detecting small high-$z$ galaxies as described in \cite{Coe2019}. \textit{HST} reduced images and catalogues are publicly available on Mikulski Archive for Space Telescopes (MAST\footnote{\url{https://archive.stsci.edu/prepds/relics/}})

\begin{figure*}
\centering
\includegraphics[width=.7\textwidth]{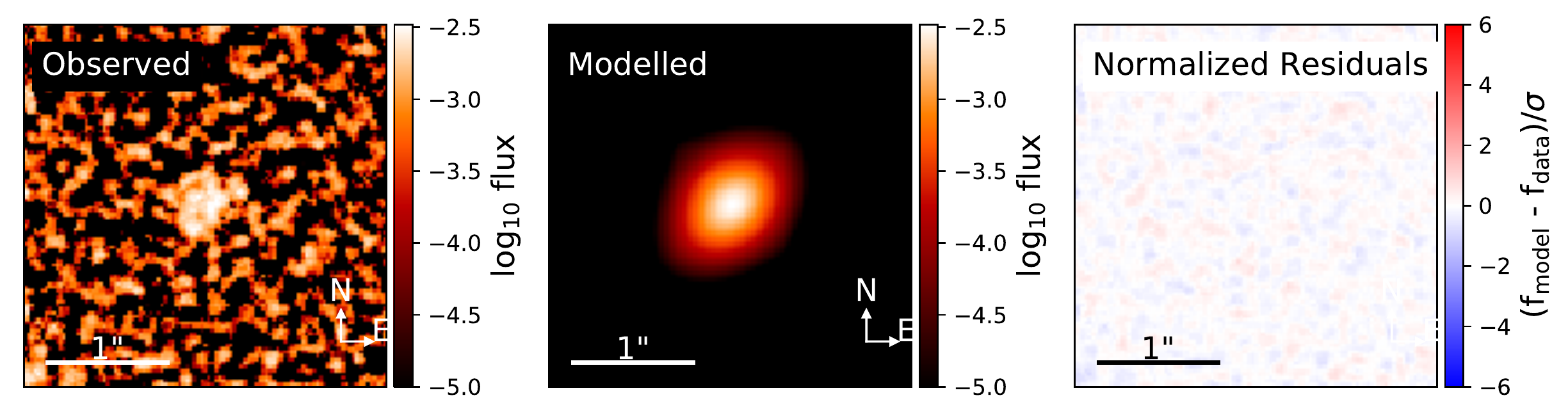}
\includegraphics[width=.25\textwidth]{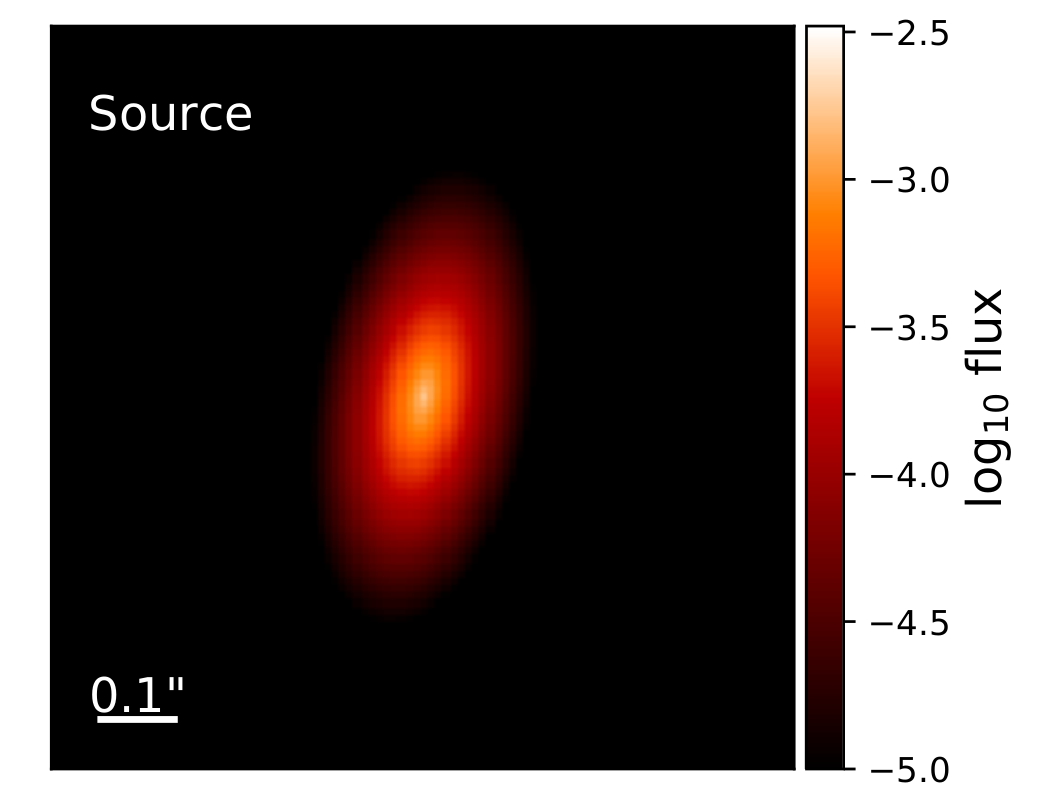}
\caption{Modeling results of the source Abell1763-1564 using \textsc{Lenstruction}, including the observed (HST/WFC3 F160W) and modelled source in the image plane with normalised residuals and the predicted morphology in the source plane (note that the source plane image is plotted on a different spatial scale), respectively from left to right. The measured size of Abell1763-1564 is $R_{\rm{eff,circ}} = 0.515$ kpc.}
\label{fig:lenstruction}
\end{figure*}

The targets in this work were also observed with \textit{Spitzer} Space Telescope via the \textit{Spitzer}-RELICS program (S-RELICS, PI Brada{\v c} \#12005, 13165, 13210, PI Soifer, \#12123), which collected $>1000$ hours of data in the 3.6$\mu$m and 4.5$\mu$m IRAC bands. 13 clusters were observed to 30-hour depth per \textit{Spitzer}/IRAC band (3-$\sigma$ depth $\sim$26 mag), using a combination of S-RELICS and archival data. The remaining 27 clusters were observed to 5-hour depth (3-$\sigma$ depth $\sim$24 mag). The details of \textit{Spitzer} data collection and reduction, as well as a complete catalogue of \textit{HST}+\textit{Spitzer} fluxes and a variety of stellar properties for each high-redshift object can be found in \cite{Strait2020}. Reduced \textit{Spitzer} images are available from the NASA/IPAC Infrared Science Archive (IRSA\footnote{\url{https://irsa.ipac.caltech.edu/data/SPITZER/SRELICS/}}) and all raw data are available for download on the \textit{Spitzer} Heritage Archive (SHA\footnote{\url{https://sha.ipac.caltech.edu/applications/Spitzer/SHA/}}).

\subsection{Lens Models}\label{lensmodels}
In order to model the high-redshift galaxies in their source planes, we utilized magnification maps primarily from Glafic \citep{Oguri2010} lens models since these maps were avilable for most clusters, and we used Lenstool \citep{Jullo2009} maps for the remaining clusters. These maps were provided by the RELICS team and are made available for public use on MAST\footnote{\url{https://archive.stsci.edu/}}. Because some RELICS clusters lack sufficient constraints to create a lens model, nine of the RELICS clusters (A1300, A520, A665, PLCKG004-19, PLCKG138-10, PLCKG308-20, RXC1514-15, SPT0254-58, WHL0137-8) have no publicly available models, and thus any sources lensed by these clusters are discarded for lack of reliable magnification measurements. For further information regarding these lens models, see \citealp{Cerny2018}, \citealp{Acebron2018}, \citealp{Cibirka2018}, \citealp{PaternoMahler2018}, \citealp{Acebron2019a}, \citealp{Matthee2019}, \citealp{Acebron2019b}, \citealp{Okabe2020}. It is important to note that size measurements in the presence of lensing can be recovered robustly given an appropriate lens model, as shown by, e.g., \cite{Yang2020a}. Further, we will discuss uncertainties due to magnification in detail in Section~\ref{biasandunc}.

\subsection{Sample Selection}\label{sample}
We obtain our sample of galaxies from the RELICS catalogue \citep{Strait2020}, selecting 78 singly imaged high-redshift sources with peak redshift $z\geq5.5$. We use the photometric redshifts from Method A in \cite{Strait2020}. Briefly, these are calculated using a set of seven templates from EA$z$Y (\citealp{Brammer2008}) in linear combination to estimate the redshift. The average uncertainty on redshift is $\sim2\%$, calculated from the 68\% confidence intervals of the photo-$z$ probability density function (PDF).

From the 207 sources in the \cite{Strait2020} catalogue, we include only sources with reliable \emph{Spitzer}/IRAC fluxes for accurate spectral energy distribution (SED) fitting, which allows us to {more reliably} determine stellar masses and star formation rate (SFR).  Our sample consists of blue, star-forming galaxies selected using a variety of photometric redshift codes \citep{Salmon2020,Strait2020}. We limit our study of source sizes to galaxies detected in \textit{HST} F160W with high signal-to-noise ratio (S/N$>$5) and resolved in the image plane given the \textit{HST} point-spread-function (PSF)\footnote{\url{https://www.stsci.edu/hst/instrumentation/wfc3/data-analysis/psf}} for the Wide Field Camera 3 (WFC3)  \citep{Mackenty2008}. We also require the use of reliable lens models to constrain the properties of singly imaged sources as explained above. After these selections, our sample consists of 78 sources from 28 independent lines of sight. 

\section{Galaxy Size Measurements}\label{sizes}

To measure the sizes of sources in RELICS clusters, we closely follow the methods of \cite{Yang2020a}, utilizing the Python package \textsc{Lenstruction} \cite{Yang2020}, which is based on \textsc{Lenstronomy} (\citealp{Birrer2015}; \citealp{Birrer2018}). \textsc{Lenstruction} has been proven reliable for source reconstruction and size measurements as tested in \cite{Yang2020a}. Throughout this section, we outline the basics of the modeling procedures and uncertainties associated with our assumptions. For more information on the process, we refer the reader to \cite{Yang2020} as well as the GitHub repositories for \textsc{Lenstruction}\footnote{\url{https://github.com/ylilan/lenstruction}} and \textsc{Lenstronomy}\footnote{\url{https://github.com/sibirrer/lenstronomy}}.

\textsc{Lenstruction} adopts forward modeling, using the appearance of our high redshift galaxies in the image plane to model the morphology in the source plane. Because our sample deals strictly with high-redshift objects, we use F160W images at 30 milliarcsecond resolution for our size measurements, corresponding to $\sim1.54 \rm{\mu m}$~/~$(1 + z_{\rm{median}})\approx 0.21 \rm{\mu m} $ in rest-frame light, where $z_{\rm{median}}=6.2$ is the median redshift of the sample. The measurement thus varies slightly in rest-frame wavelengths in our sample, as we cover a range of redshifts ($5.5\lesssim z \lesssim 10$), but always remains in the rest-frame UV. We select a stellar object in the corresponding cluster's field to take into account the PSF of the instrumentation, and the modeling process also makes use of lens models (Section \ref{lensmodels}) for consideration of the distortion of each singly-imaged galaxy. We use RMS maps of each cluster made available by RELICS as inputs for the pixel selection of each object and scale these maps by 1.1 to account for correlated noise (see \citealp{Trenti2011}), although we note that this correction has no significant influences on our results. \textsc{Lenstruction} also removes the background.

We parameterize the light profile of each object's shape as an elliptical S\'{e}rsic profile, with several free parameters including the S\'{e}rsic radius, which is the semi-major half light, or effective, radius ($R_{\rm{eff}}$). We fix the S\'{e}rsic index at $n_{\rm{Sersic}}$=1 and constrain the ellipticity with an upper limit of $0.9$ in agreement with the methods of \cite{Kawamata2018}. The source parameters of the best fit solution produced by \textsc{Lenstruction} provide information about the delensed galaxy from which we measure the effective radius.

Outputs from \textsc{Lenstruction} for an example galaxy in our sample are shown in Figure \ref{fig:lenstruction}, with observations, the modeled galaxy in the image plane, a residual map, and the predicted morphology of the galaxy in the source plane.

\subsection{Biases and Uncertainties}\label{biasandunc}
\begin{figure*}
\centering
\includegraphics[width=\textwidth]{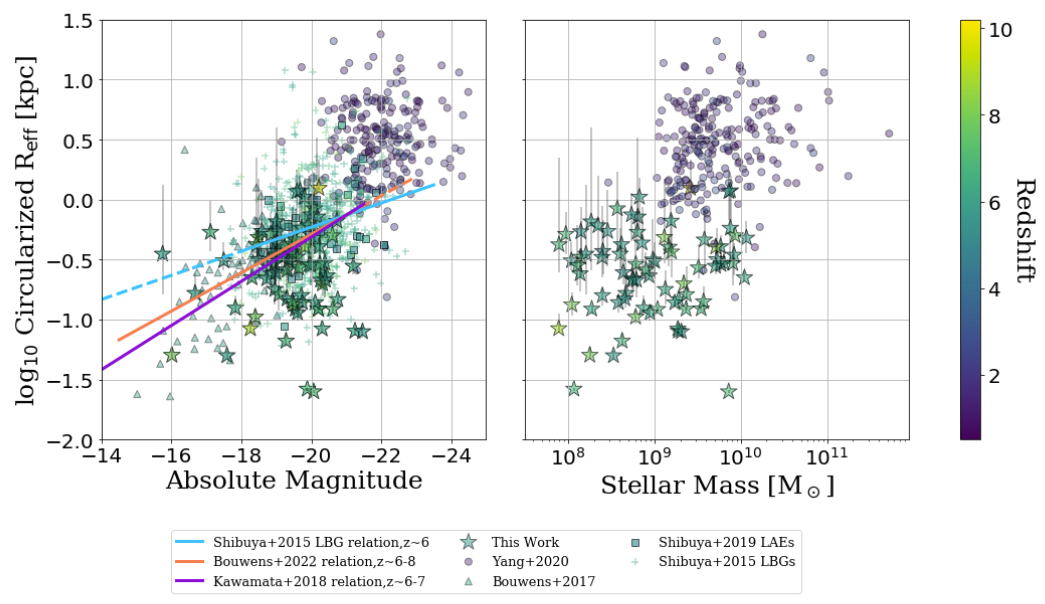}
\caption{Measured circularised effective radius vs. absolute magnitude (left) and stellar mass (right), with points coloured by redshift. Starred points indicate our data, alongside circles for $z \sim 1-3$ data from \protect\cite{Yang2020a}, triangles for {\protect\cite{Bouwens2017}}, and squares (Lyman $\alpha$ emitters) and '$+$' points (Lyman break galaxies) for \protect\cite{Shibuya2015,Shibuya2019}. {Size-magnitude relations derived by \protect\cite{Kawamata2018}, \protect\cite{Bouwens2021}, and \protect\cite{Shibuya2015} are shown as purple, orange, and cyan lines respectively on the first plot}. Errorbars on size measurements are derived from the MCMC process of Lenstruction described in Section~\ref{biasandunc}. We measure small galaxies, similar in size to the other studies and in some cases, brighter.}
\label{fig:size vs mag}
\end{figure*}

The statistical uncertainties on size measurements are obtained from a Markov chain Monte Carlo (MCMC) process. This process samples over parameters $R_{\rm{eff}}$, eccentricity, and spatial position of the galaxy in the source plane to investigate the confidence intervals of the model and degeneracies among parameters. Results are characterized by a 1$\sigma$ distribution, and errors are included in Figure \ref{fig:size vs mag}, with an average uncertainty on $R_{\rm{eff}}$ of 53.4\%. 

One possible bias is induced by our assumption of a S\'{e}rsic index of 1. While $n=1$ is perhaps a reasonable assumption for $z\geq5.5$ galaxies (\citealp{Petty2009}, \citealp{Buitrago2013}, \citealp{Krywult2017}, \citealp{Bouwens2021}), our data do not have the power to constrain S\'{e}rsic indices and therefore this choice may introduce a systematic error in our measurements of effective radii. If we assume a S\'{e}rsic index of $n=0.5$ or $n=2$, our half-light radius results differ by an average of less than 6\%, and this assumption is thus negligible on the average uncertainty. 

To estimate other systematic uncertainties, we apply a series of tests as described below. In our analysis, we assume the peak redshift from the PDF of redshift distribution, calculated from the spectral energy distribution (SED) fitting method described in \citeauthor{Strait2020}(\citeyear{Strait2020}, Method A), as well as best fit magnification from lens models (described in Section \ref{lensmodels}). There is an uncertainty in both redshift and magnification for each source. We estimate the propagation of these uncertainties on sizes by obtaining the sizes of the sources using the convergence and shear maps from the range of maps calculated by bootstrapping lens modeling constraints, which incorporates those statistical uncertainties. In addition, we randomly select a redshift from the galaxy's photo-$z$ PDF for each iteration to estimate the effect of the uncertain redshift on our results. We then calculate 68\% confidence intervals from the combined magnification and redshift uncertainties and find that these errors contribute less than an additional 10\% to size measurement errors. Because these uncertainties are much smaller than those reported from \textsc{Lenstruction} due to uncertainties from lensing itself, the uncertainties from S\'{e}rsic index and magnification and redshift are not included in Figure \ref{fig:size vs mag}.

To calculate the cosmic variance of one RELICS field we use the CosmicVarianceCalculator\footnote{\url{https://www.ph.unimelb.edu.au/~mtrenti/cvc/CosmicVariance.html}} (\citealp{Trenti2008}), and using inputs of mean redshift and number of objects detected per field, we get a value of 0.3. The total variance reduces by a factor of $\sqrt{N}$, with $N$ being the 28 independent fields our 78 candidates are found within, giving 5.7\% as the fractional error on galaxy number counts due to sample variance. 

In addition to these uncertainties, we calculate the completeness of our sample to understand our results relative to the what the galaxy population likely is. Though lensing is an incredibly useful tool, understanding incompleteness in lensing fields is more complicated than doing so in blank fields. Specifically, completeness is a function not just of galaxy luminosity and size, but also of magnification and distortion, and thus varies with position with respect to the lensing cluster and the lensing properties of the lensing cluster (e.g., \citealp{Bouwens2021}). 

{Using the software GLACiAR2\footnote{\url{https://github.com/nleethochawalit/GLACiAR2-master}} (\citealp{Carrasco2018,Leet2022}), we calculate source detection completeness as a function of magnitude and size for a typical cluster observation in our sample. We follow the methods in \cite{Yang2022}, with the code modified to incorporate effects from gravitational lensing by utilizing available lens models. This software simulates lensed galaxies modeled with a S\'{e}rsic profile at random positions in the source plane and predicts their lensed properties in the image plane, calculating the detection recovery fraction in the F160W image. We set bins for absolute magnitude, $M_\text{UV}$, from -22 to -16 and sizes, $R_\text{eff}$, from $0.03$ to $1$ kpc, as well as redshifts $z\sim5.5-10$, which are characteristic ranges of our sample. For more details on the process, refer to \cite{Yang2022}.}

\begin{figure}
\centering
\includegraphics[width=\linewidth]{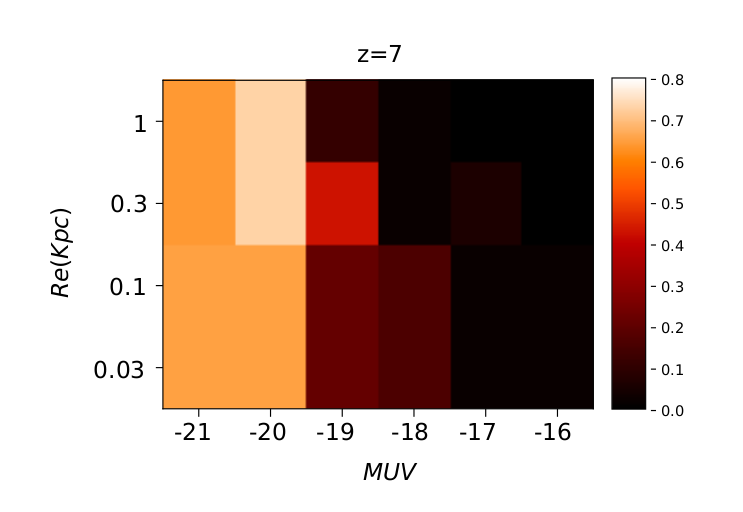}
\caption{{Detection completeness for a typical cluster observation in our sample (Abell 1763). We simulate $z\sim7$ galaxies with different magnitudes and sizes and measure their recovery fraction (color bar). Our sample is relatively complete ($\geq50\%$) for brighter galaxies at magnitudes of -19 to -21 and small sizes $<0.3$ kpc, but our sample might be missing large and/or faint galaxies and is very incomplete for the faintest population. Here we show completeness detections for only one redshift ($z=7$) because the redshift dependence within our redshift range is negligible for completeness calculations.}}
\label{fig:completness}
\end{figure}

{An example of the resulting completeness map for the cluster Abell 1763 at a redshift of $z=7$ is shown in Figure~\ref{fig:completness}. For sources between -19 and -21 $M_{\rm{UV}}$, completeness is about 60-80\%. While we are relatively complete for this brighter population of galaxies, our sample is incomplete for the faintest population ($<10$\%) and we may be preferentially missing large, faint galaxies.}

We make no attempt in this work to constrain the entire population of high redshift galaxies and rather demonstrate that a population of compact galaxies at $z\geq 5.5$ does exist. With \textit{JWST}, we will improve on incompleteness measurements at high redshift. This goal of this work is to produce an exploration of the distribution of sizes in a sample of star-forming high redshift galaxies rather than to reconstruct the intrinsic size-luminosity and size-mass relations. Therefore we leave a more complete characterization of the population to future work; however, this important caveat should be kept in mind when interpreting our results.

\section{Results and Discussion}\label{results}

\subsection{Small Sizes}\label{smallsizes}
Size measurements and other properties of selected galaxies from our sample are provided in Table \ref{table:1}, and we present our results for the size-magnitude and size-mass measurements of our sample in Figure \ref{fig:size vs mag}. We analyse our results in the context of previous studies, using the circularised effective radii of our sources ($R_{\rm{eff,circ}} = R_{\rm{maj}} * \sqrt{1 - e}$, where $R_{\rm{maj}}$ is our measured effective radius, $e =1 - \sqrt{1 - c^2}$ is ellipticity, and $c$ is eccentricity as determined by the best fit parameters output by Lenstruction). The circularized effective radius is the standard parameter for the size measurements of small, high redshift sources (e.g, \citealp{Ono2013,Holwerda2015,Shibuya2015,Shibuya2019}). On the left of Figure \ref{fig:size vs mag} we plot measured sizes against magnitudes, where starred points are our data, triangles are data from {\cite{Bouwens2017}} ({also used in \citealp{Bouwens2021}}) who uses deeper data, but in fewer lensed fields, and circles are low redshift data from \cite{Yang2020a}, and each point is coloured by redshift. Our sample contains brighter galaxies compared to the fainter population from {\cite{Bouwens2017}}, and both measure small sizes ($<200$ pc). We also include sizes of $z\geq 6$ Lyman-$\alpha$ emitters (LAEs) and Lyman break galaxies (LBGs) from \cite{Shibuya2015,Shibuya2019} as squares and plus signs, respectively. While \cite{Shibuya2019} have a much larger sample of star-forming galaxies, we probe smaller sizes due to lensing.

We also compare our size measurements to previous studies in the context of the size-mass relation of galaxies in Figure \ref{fig:size vs mag} on the right. {We use median stellar mass measurements as obtained by \citeauthor{Strait2020} (\citeyear{Strait2020}, Method A), estimated by the median of the probability density function (PDF) of each source.} Once again, our data is shown as starred points along with data from \cite{Yang2020a} as circles, coloured by redshift. 

The presence of small galaxies is consistent with the results of {\cite{Bouwens2017,Bouwens2021}} and \cite{Kawamata2015}, with galaxy sizes on the order of $\sim30-100$ pc, as well as the small sizes found by \cite{Coe2013} and \cite{Zitrin2014} at the highest redshifts.  {We include size-luminosity relations derived by \cite{Shibuya2015}, \cite{Bouwens2021}, and \cite{Kawamata2018} as lines in Figure~\ref{fig:size vs mag}, from which we can see that} these small sizes ($R_{\rm{eff}} < 200$ pc) result in a steep size-luminosity relation, although a characterization of the incompleteness and selection function are needed to quantify this trend. While our sample is incomplete, these results show that compact star forming galaxies, only slightly larger than typical globular clusters, exist in non-negligible numbers in the early universe. 

\subsection{Implications for Reionization}\label{implications}

We can use our size measurements to compare to other properties of these high-redshift galaxies as well. Through the investigation of these properties, we can gain further insight into the process of reionisation in the early universe and the role of these high-redshift galaxies as possible Lyman continuum leakers.

\subsubsection{Star Formation Rate Surface Density and [OIII]+H$\beta$}\label{sfrd results}

Previous works have used compact sizes to select for potential Lyman continuum leaking galaxies, {including the compact Green Pea galaxies at $z\sim0.1-0.3$, which are likely relevant local analogs of high-redshift UV galaxies \citep{cardamone2009,Kim2021}. Lyman continuum leakers are also thought to be compact and highly star-forming. For example, \cite{Marchi2018} find that UV compact sources ($R_\text{eff}<0.3$ kpc) at $z\sim3.5-4.3$ with high SFR surface density ($\Sigma_\text{SFR}$) are likely to have higher Lyman continuum flux than lower surface density galaxies. Additionally, \cite{Izotov2016b} find that a sample of $z\sim0.3$ known Lyman continuum leakers have high SFR surface densities. Other works also find a correlation between Lyman continuum escape fraction and $\Sigma_\text{SFR}$ in low-redshift analogs at $z\sim0.2-0.4$ \citep{Flury2022} and $z\sim0.3$ \citep{Verhamme2017}. Using the SFR surface densities of known Lyman continuum galaxies, \cite{Naidu2020} find that Lyman continuum leakers have higher $\Sigma_\text{SFR}$ than average galaxies at their respective redshifts.} In addition, it has been found that the velocity of ionised outflows is strongly correlated with SFR and $\Sigma_\text{SFR}$ (e.g., \citealp{Heckman2015,Cicone2016}). {While only known to be applicable for lower redshift galaxies that may not be the exact analogs of the high-redshift universe, this is currently one of the best criteria we have to indirectly identify potential Lyman continuum leakers.} 

\begin{figure}
\centering
\includegraphics[width=\linewidth]{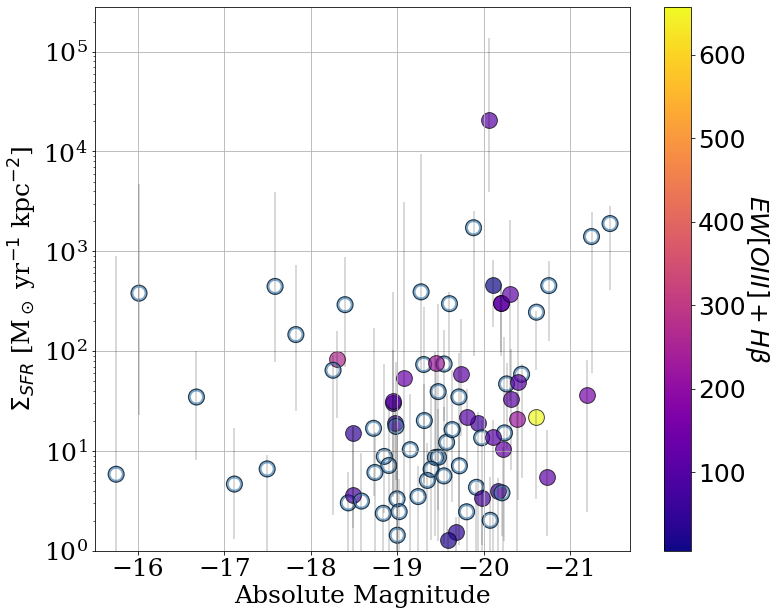}
\caption{Star formation surface density, calculated using our measurements of circularised effective radius and SFR from \protect\cite{Strait2021}, vs. absolute magnitude. {There are many sources with SFR surface density $> 10$ $M_\odot$ yr$^{-1}$, which is higher than average for typical galaxies at this redshift {increasing the likelihood that these} are potential Lyman continuum leakers. Sources are colored based on the colorbar on the right, {indicating the values of the [OIII] + H$\beta$ EW} for detections (S/N$>$3) in channels 1 or 2 of IRAC.}}\label{fig:sfrd vs mag}
\end{figure}

{We calculate the SFR surface density using the relation $\Sigma_{\rm{SFR}} = \rm{SFR}/ (2\pi R_\text{eff}^2)$ \citep{Ono2013,Shibuya2019,Naidu2020,Flury2022}, where SFR is the star formation rate as calculated by \citeauthor{Strait2020} (\citeyear{Strait2020}, Method A) and $R_\text{eff}$ is our measured circularised effective radius of the source.} We present the distribution of our sample in the SFR surface density-absolute magnitude plane in Figure \ref{fig:sfrd vs mag}. {As seen in this figure, our sources have high SFR surface densities, as they are at relatively high redshifts and the RELICS survey tends to select relatively bright sources in the UV which generally scales with SFR \citep{Kennicutt1998,Lemaux2021}. We also detect some intrinsically faint sources due to lensing. There are several sources with $\Sigma_\text{SFR}>10 M_\odot \text{yr}^{-1} \text{(kpc)}^{-2}$, which is higher than the average for typical galaxies at redshifts $z\sim6-8$ according to the relation between $\Sigma_\text{SFR}$ and redshift found by \cite{Naidu2020} and \cite{Shibuya2015}. With high SFR surface densities, these galaxies are potential Lyman continuum leaking candidates.}

In addition, a common indicator of galaxies with highly ionised environments is strong H$\beta$ $\lambda$4861\AA\ + [OIII] $\lambda$4959,5007\AA\ emission. This indicator has been used in previous works (e.g., \citealp{Verhamme2015,Izotov2016b,Fletcher2019,Izotov2021,Malkan2021}) to select Lyman continuum leakers at lower redshifts, and {\cite{tang2019} finds that ionizing photon production increases with higher [OIII] + H$\beta$ EW at $z=1.3-2.4$. Several studies find that high-redshift ($z\gtrsim7$) galaxies typically have extreme [OIII]+H$\beta$ emission, with equivalent widths greater than 1000\AA~(e.g., \citealp{Smit2015,borsani2016,borsani2020}).  \cite{Endsley2021} also finds that extreme emitters at $z\sim7$ with a median [OIII] + H$\beta$ EW of $\sim$ 700 \AA ~may be undergoing a short burst of star formation, which increases the escape fraction of ionizing photons.}. While \cite{Saxena2021} finds that there is no strong correlation between the [OIII]+H$\beta$ emission of a galaxy alone and its Lyman continuum radiation escape fraction $f_{\rm{esc}}$, [OIII] + H$\beta$ emission is likely a useful tool to select possible Lyman continuum leakers in conjunction with other probes such as $\Sigma_{\rm{SFR}}$. {We plot [OIII] + H$\beta$ equivalent widths as a colorbar in Figure \ref{fig:sfrd vs mag} for sources with IRAC channel 1 or 2 detections (S/N$>3$) as measured by \citeauthor{Strait2020} (\citeyear{Strait2020}, Method B). Because [OIII]+H$\beta$ is not well {constrained for all objects in our sample}, and there is uncertainty on {the use of the [OIII]+H$\beta$ EW} as a criterion for Lyman continuum leakages, we do not emphasize it as a criterion for our sample.
}

\begin{figure*}
\centering
\includegraphics[width=.3\textwidth]{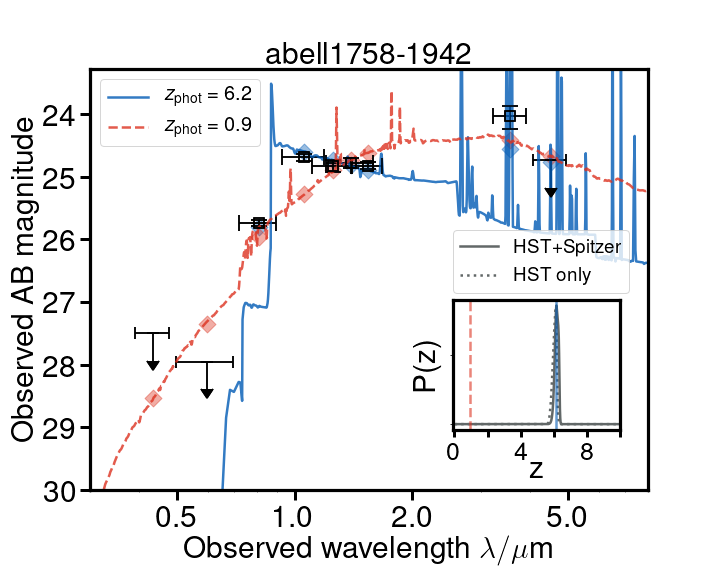}
\includegraphics[width=.3\textwidth]{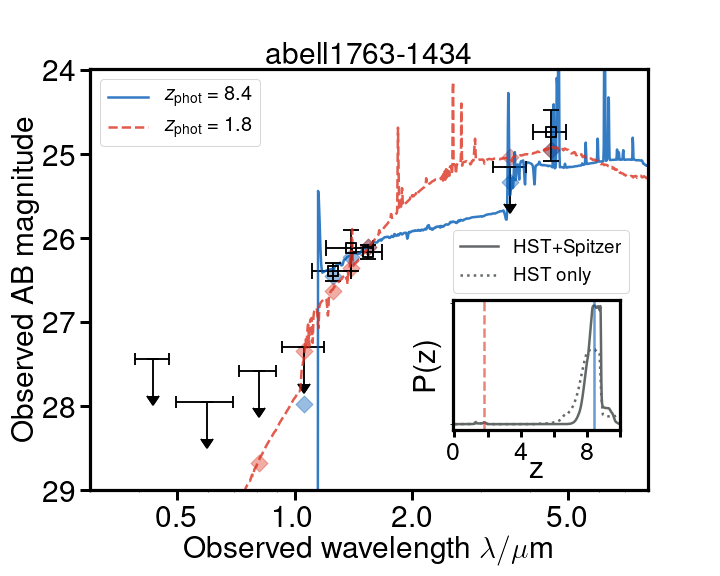}
\includegraphics[width=.3\textwidth]{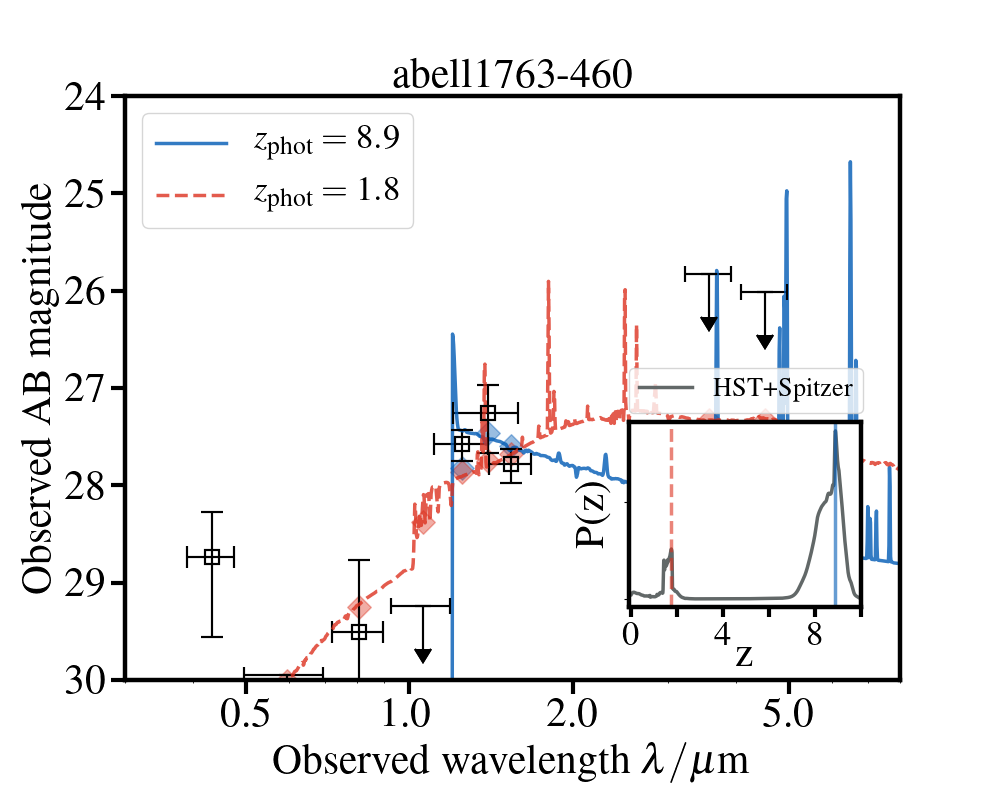}
\includegraphics[width=.3\textwidth]{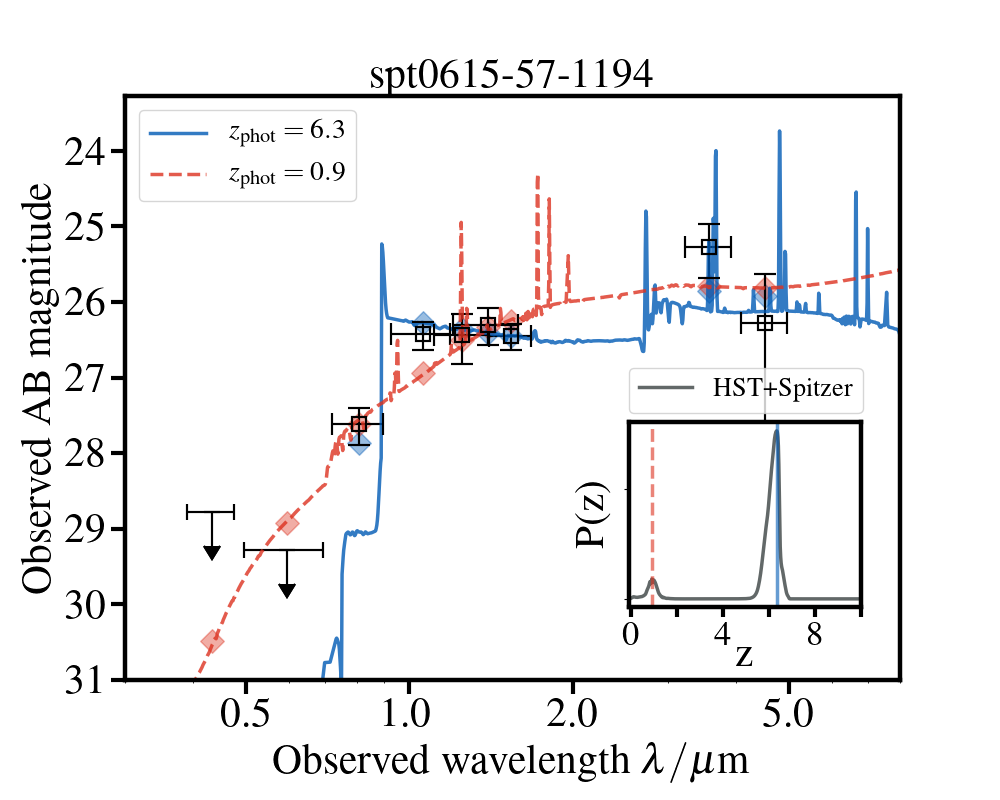}
\includegraphics[width=.3\textwidth]{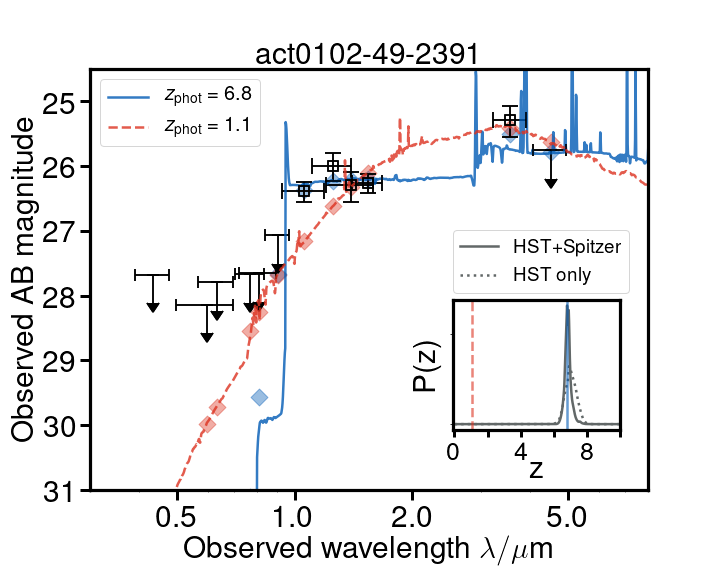}
\caption{Spectral energy distributions for {14 examples of galaxies with high SFRD and high inferred [OIII]+H$\beta$ emission, obtained using Method A as described by \protect\cite{Strait2020}. Black points are data, and blue and red lines are predicted photometry from models using the best-fit high redshift and secondary redshift peak templates, respectively, with the redshift PDF for each source shown in the insets. Each of these galaxies are at similar redshifts ($z\sim 7$), where [OIII]+H$\beta$ fall into the channel 1 filter of Spitzer/IRAC.} When extreme in emission, the flux in this filter is boosted. Boosted nebular emission and high SFRD are two qualities that tend to yield Lyman continuum leaking galaxies. It is unknown whether these galaxies are Ly$\alpha$ emitters due to the lack of spectroscopic data of sufficient depth.}
\label{fig:sed}
\end{figure*}

{
{We present the best-fit {SEDs} for a selection of these potential Lyman continuum leaking galaxies, with high SFR surface densities and inferred [OIII]+H$\beta$ EW, in Figure~\ref{fig:sed}}. Several of these galaxies show excess emission in channel 1 of \textit{Spitzer}/IRAC, implying a strong presence of nebular emission. These include the extreme [OIII] emitter, PLCGK287+32-2013 with [OIII] + H$\beta$ EW$\sim 700$\AA, highlighted in \cite{Strait2021} and the quadruply imaged galaxy in \cite{Zitrin2017}.}

\subsubsection{Lyman $\alpha$ Emission}\label{lya}

High [OIII] + H$\beta$ emission likely implies that a galaxy's interstellar medium (ISM) is more transparent to Lyman continuum photons and points to high specific star formation rate, active galactic nuclei (AGN), or both, which is correlated with the strength of outflows (\citealp{Chisholm2017}). Ly$\alpha$ emission is another possible probe of the capability of the galaxy to output ionising photons into the surrounding IGM (e.g., \citealp{Shapley2016}). If Ly$\alpha$ is visible at $z>6$, there must be either a large enough ionised bubble for the Ly$\alpha$ photons to escape or an outflow and leaking Lyman continuum photons. When observed together, strong Ly$\alpha$ and [OIII]+H$\beta$ emission thus indicate a high likelihood for a galaxy to be a Lyman continuum leaker. 

The majority of galaxies in this sample do not have sufficiently deep spectral data to have observed Ly$\alpha$ emission with the exception of MS1008-12-427. This is the Dichromatic Primeval Galaxy at $z\sim7$ (DP7) discovered by \cite{Pelliccia2021}. With extreme Ly$\alpha$ emission ($>200$\AA\ \thinspace equivalent width)  {compared to other LAEs at $z \sim 7$ (e.g., \citealp{pentericci2014,hoag2019b,Fuller2020})}, it has the potential to be a Lyman continuum leaker, or to be residing in an ionised bubble nearby a Lyman continuum leaker. \cite{Pelliccia2021} describe the possibility of DP7 being host to two distinct components, one dusty and red, and the other highly star forming and blue, and this possible merger or rejuvination event could lead to a large outflow, producing the strong Ly$\alpha$ emission. {The very red SED and distinct UV $\beta$ slopes suggest that a majority of the UV flux comes from the blue component. We measure the combined effective radius of both components, as the sources are blended, but the possibility of two components suggests that the measured size of this source may be that of two distinct objects. Reducing the effective radius by half and attributing all of the UV flux to the bluer component results in double the star formation rate surface density.} {With $\Sigma_\text{SFR}\sim11^{+55}_{-2} M_\odot$ yr$^{-1}$ (kpc)$^{-2}$, the SFR surface density of DP7 then falls within our selection criterion.}

PLCKG287-2013 and the quadruply imaged galaxy 777 have been observed with GMOS Gemini South but no Ly$\alpha$ emission was detected (see \citealp{Strait2021}, Section 5.4.2 for further details). Regarding the rest of the sample, we can also compare with LAEs from \cite{Shibuya2019} in Figure \ref{fig:size vs mag} to infer where LAEs may lie for our sample in the size-magnitude space. If any of these galaxies are Ly$\alpha$ emitters, they could be important sources for understanding the output of ionising photons in the Epoch of Reionisation, as they are compact and highly star-forming. In addition, these sub-L* galaxies are especially important for follow up because, while they are not as bright as galaxies quoted by \citep{Naidu2021}, they can provide further understanding of the contribution of a range of "medium-bright" representative galaxies to reionisation.

\begin{table*}
\caption{{Size measurements and other selected properties of sources in our sample. Full table including all sources is available in the electronic edition as supplementary material.}}\label{table:1}
\centering
\begin{tabular}{ c c c c c c c c c}
\hline \hline
Object ID & $z_{\rm{peak}}^1$  & $M_{\rm{UV}}^2$ & M$_*^3$ & $R_{\rm{eff}}^4$ & $R_{\rm{eff,circ}}^5$ & $\Sigma_{\rm{SFR}}^6$ &$\mu^7$ & [OIII]+H$\beta$ EW $^8$\\
 & &  (mag) & ($10^9$ $M_\odot$) & (kpc) & (kpc) & ($M_\odot$ yr$^{-1}$ kpc$^{-2})$ &\\
\hline
ABELL2813-102 & 5.67  & -19.94 & 6$^{+3}_{-6}$ & 0.33$^{+0.11}_{-0.08}$ & 0.29$^{+0.12}_{-0.06}$ & 19$^{+5}_{-18}$ & 3.5 & 74$^{+2}_{-2}$\\
RXC0949+17-834 & 5.92  & -17.59 & 0.34$^{+0.13}_{-0.16}$ & 0.05$^{+0.05}_{-0.03}$ & 0.05$^{+0.04}_{-0.03}$ & 444$^{+3499}_{-366}$ & 10.3 & - \\
ABELL3192-728 & 7.00  & -18.43 & 0.08$^{+0.04}_{-0.05}$ & 0.5$^{+0.7}_{-0.3}$ & 0.4$^{+0.7}_{-0.2}$ & 3$^{+3}_{-2}$ & 3.1 & - \\
PLCKG209+10-1145 & 6.00  & -19.72 & 8$^{+3}_{-3}$ & 0.6$^{+0.4}_{-0.3}$ & 0.60$^{+0.3}_{-0.3}$ & 7$^{+11}_{-5}$ & 3.0 & - \\
PLCKG209+10-202 & 6.15  & -20.76 & 1.9$^{+1.3}_{-1.7}$ & 0.15$^{+0.04}_{-0.05}$ & 0.15$^{+0.04}_{-0.05}$ & 453$^{+345}_{-327}$ & 3.0 & - \\
PLCKG287+32-2013 & 7.54 &  -20.60 & 9$^{+5}_{-6}$ & 0.52$^{+0.02}_{-0.02}$ & 0.502$^{+0.020}_{-0.019}$ & 22$^{+3}_{-18}$ & 4.2 & 656$^{+11}_{-11}$\\
PLCKG287+32-2032 & 7.90 & -16.01 & 0.18$^{+0.06}_{-0.12}$ & 0.06$^{+0.08}_{-0.04}$ & 0.05$^{+0.07}_{-0.04}$ & 380$^{+4293}_{-357}$ & 56.0 & - \\
PLCKG287+32-2235 & 6.97  & -20.61 & 1.1$^{+3.4}_{-0.6}$ & 0.12$^{+0.023}_{-0.019}$ & 0.12$^{+0.021}_{-0.018}$ & 245$^{+6}_{-181}$ & 2.9 & - \\
PLCKG287+32-698 & 6.86  & -20.20 & 3.5$^{+0.9}_{-1.4}$ & 0.124$^{+0.023}_{-0.019}$ & 0.123$^{+0.021}_{-0.018}$ & 303$^{+167}_{-214}$ & 4.8 & 114$^{+2}_{-2}$\\
PLCKG287+32-777 & 6.93  & -20.20 & 2.3$^{+0.4}_{-0.9}$ & 0.124$^{+0.023}_{-0.019}$ & 0.123$^{+0.021}_{-0.018}$ & 302$^{+131}_{-203}$ & 4.2 & 120$^{+3}_{-3}$\\
\hline
\multicolumn{9}{l}{\small $^1$ Peak redshift as described in Section \ref{sample}. Photometric errors are described in \cite{Strait2021}. }\\
\multicolumn{9}{l}{\small $^2$ Intrinsic absolute magnitude. }\\
\multicolumn{9}{l}{\small $^3$ Stellar mass as described in Section \ref{results}.}\\
\multicolumn{9}{l}{\small $^4$ Measured S\'{e}rsic {(effective)} radius in kpc, described in Section \ref{sizes}. }\\
\multicolumn{9}{l}{\small $^5$ Circularized effective radius in kpc, described in Section \ref{results}.}\\
\multicolumn{9}{l}{\small $^6$ Star formation surface density as described in Section \ref{results}. }\\
\multicolumn{9}{l}{\small $^7$ Magnification, described in Section \ref{lensmodels}. }\\
\multicolumn{9}{l}{\small $^8$ [OIII]+H$\beta$ equivalent width {for IRAC S/N>3 detections} as measured by \citeauthor{Strait2020} (\citeyear{Strait2020}, Method B).}
\end{tabular}
\end{table*}

\begin{figure*}
\centering
\includegraphics[width=.3\textwidth]{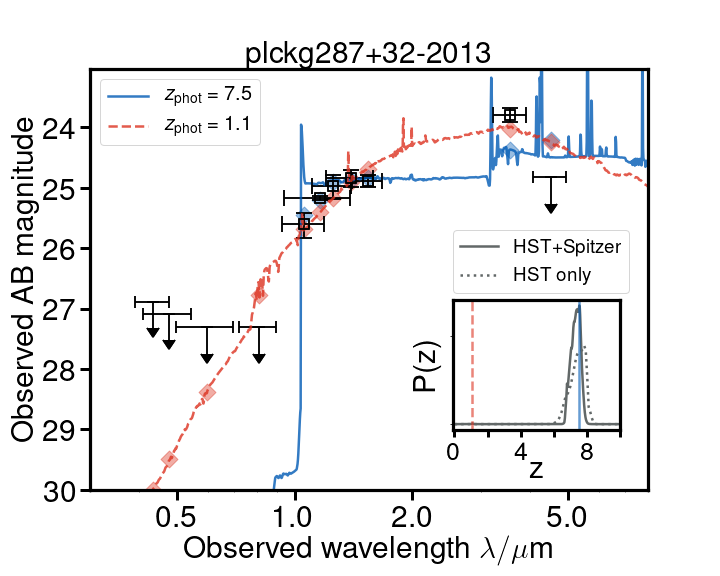}
\includegraphics[width=.3\textwidth]{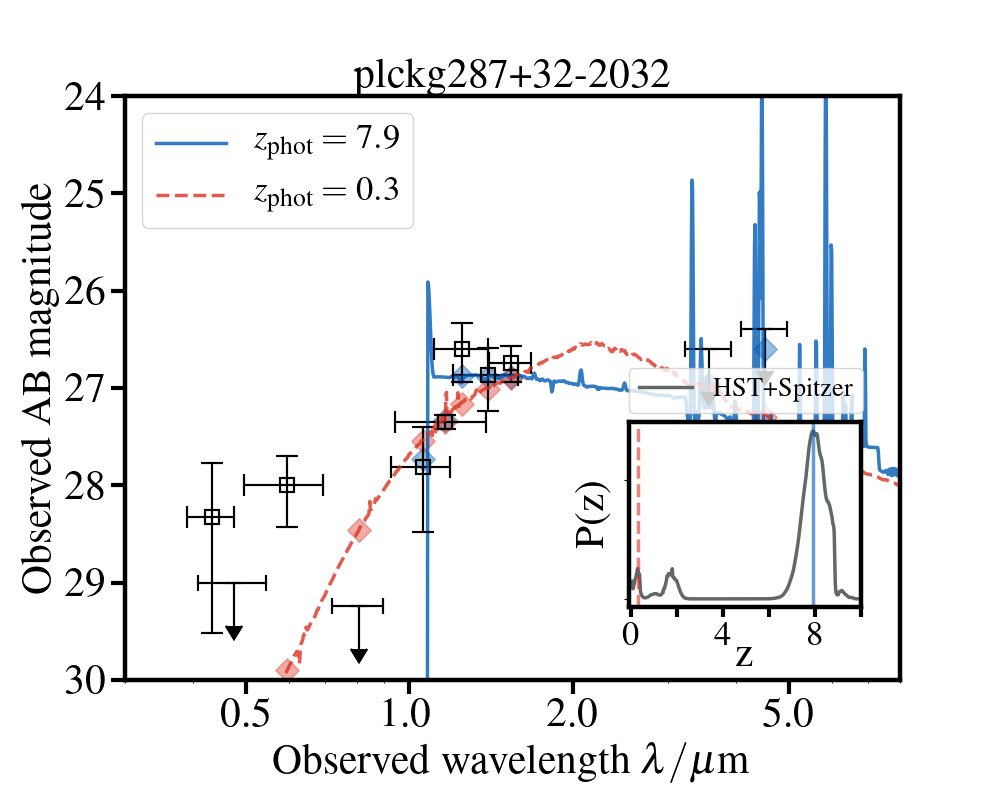}
\includegraphics[width=.3\textwidth]{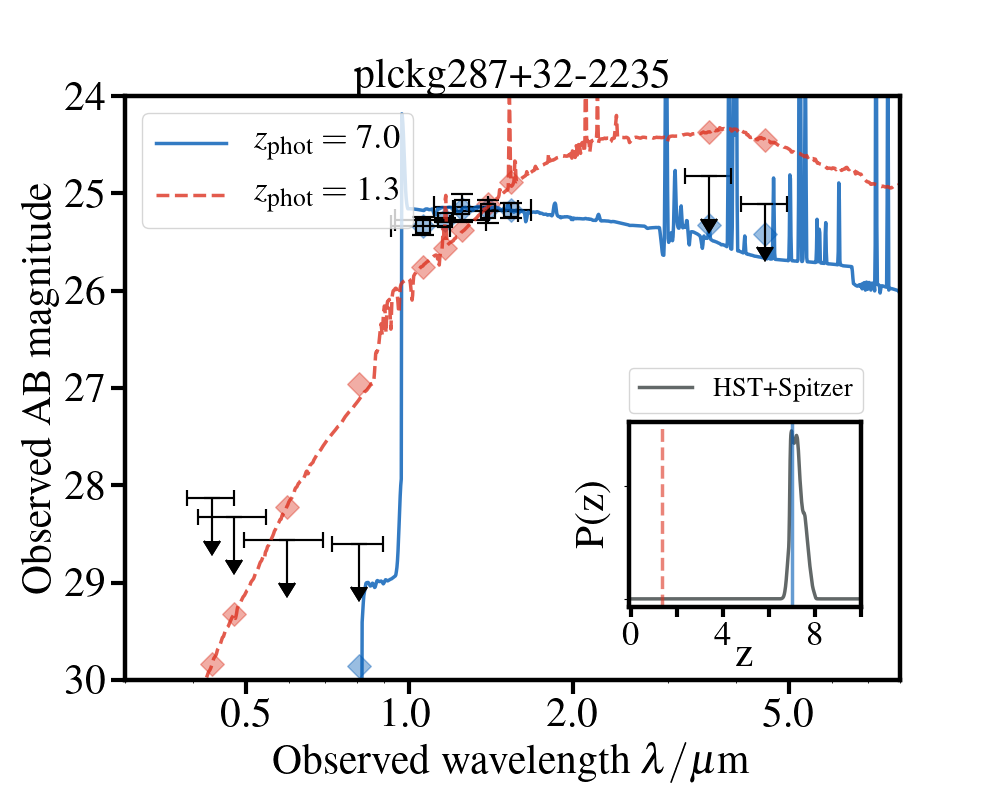}
\includegraphics[width=.3\textwidth]{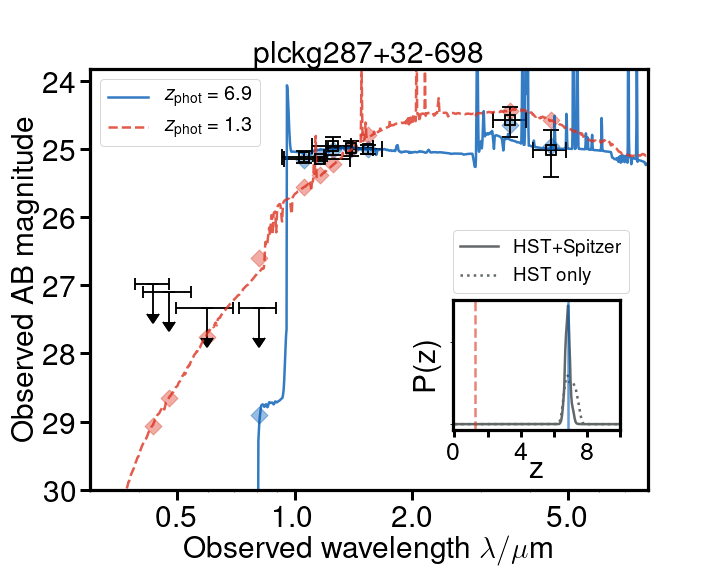}
\includegraphics[width=.3\textwidth]{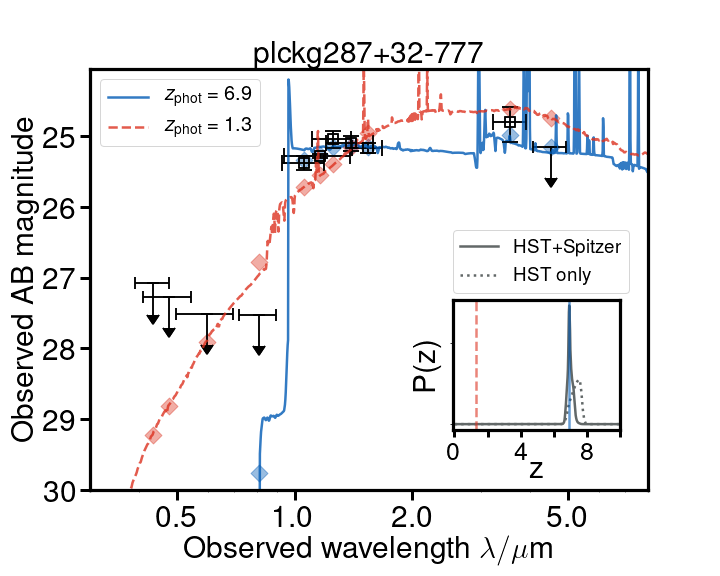}
\includegraphics[width=.3\textwidth]{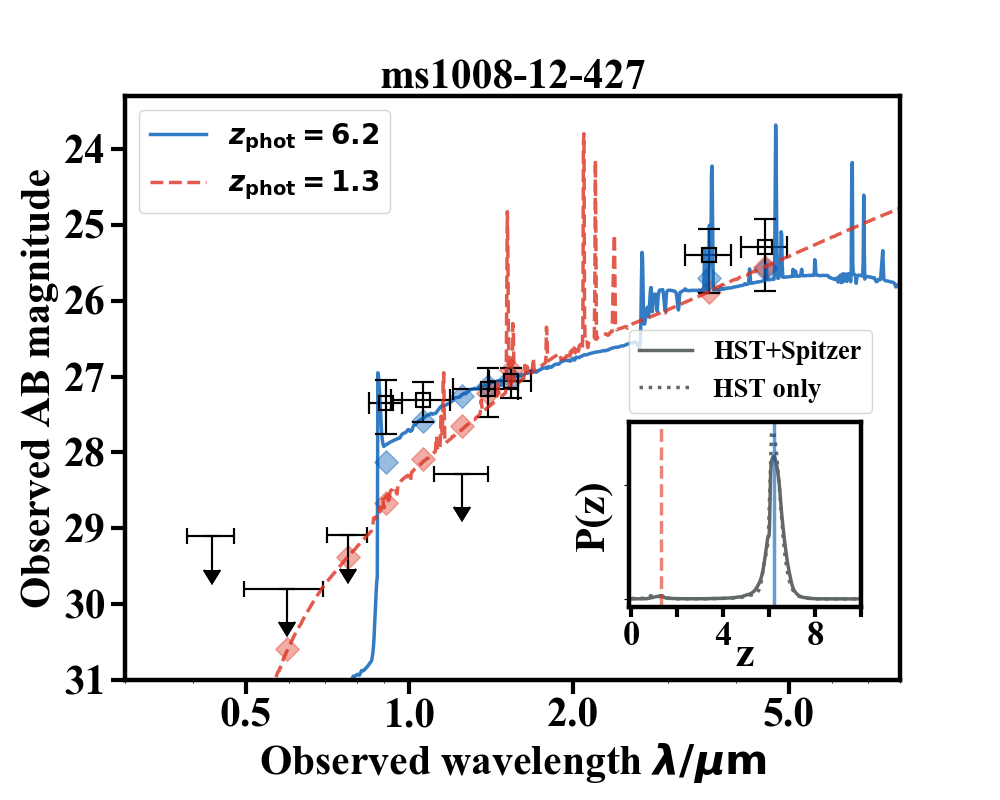}
\includegraphics[width=.3\textwidth]{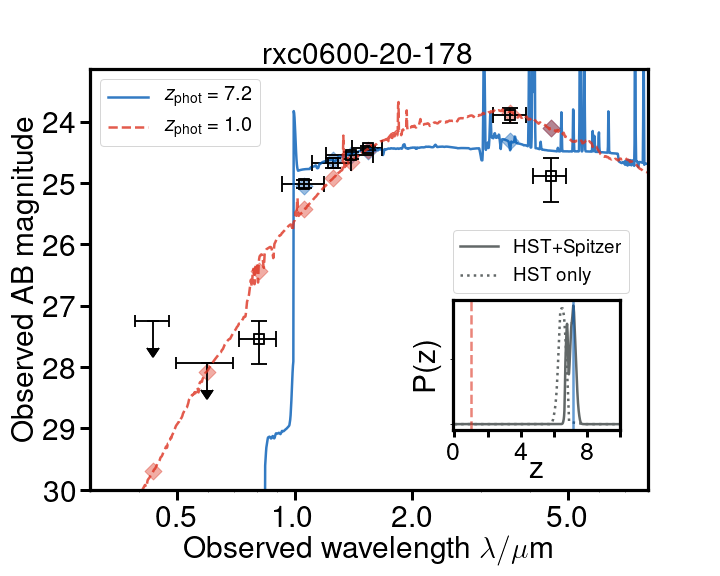}
\includegraphics[width=.3\textwidth]{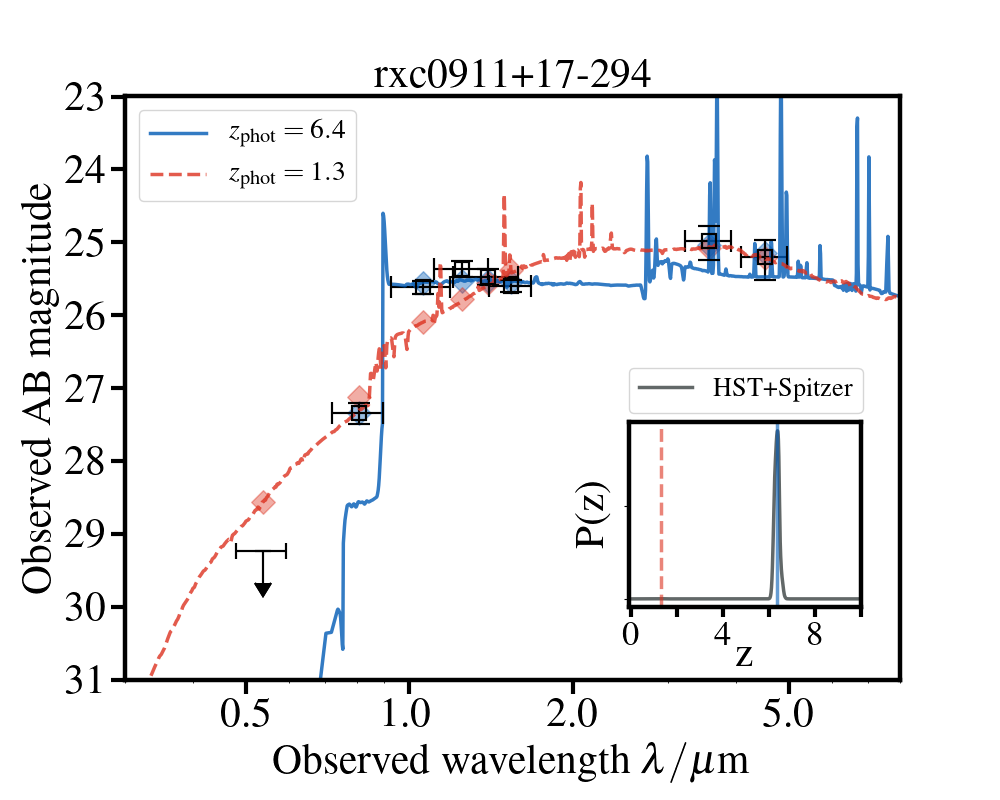}
\includegraphics[width=.3\textwidth]{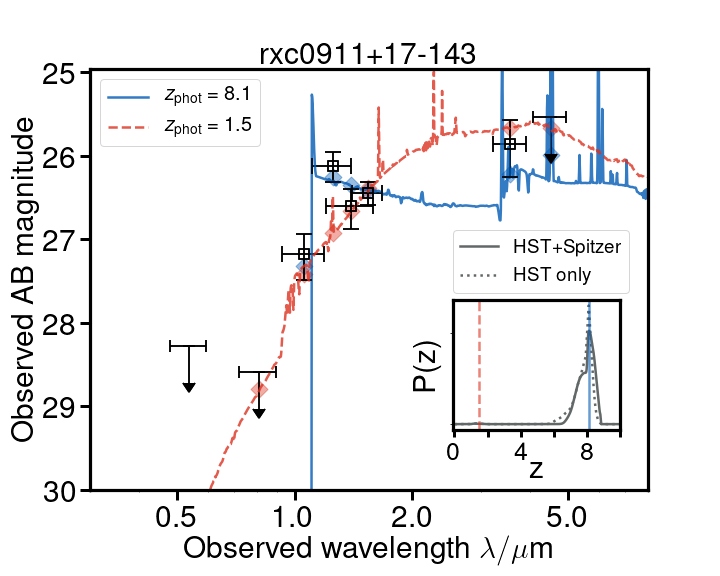}
\end{figure*}

\section{Conclusions and Future Work}\label{concls}

We present size measurements for 78 high-redshift galaxies from RELICS data, imaged by \textit{HST} and \textit{Spitzer}, utilizing the code Lenstruction to measure the sizes of these galaxies. Our main results are as follows:

\begin{itemize}
  \item We analyse the size-magnitude and size-mass measurements in the context of previous studies, finding many compact ($<$ 200 pc) galaxies in agreement with other works. These extremely small sizes seem to imply a steep size-luminosity relation, though definitive conclusions will require detailed studies to properly evaluate completeness.
  \item We calculate the star formation surface density of these sources using our measured sizes and {find several sources with higher than average SFR surface density ($\Sigma_{SFR} > 10$ $M_\odot$ yr$^{-1}$ (kpc)$^{-2}$). We highlight galaxies with high SFR surface density and inferred [OIII] + H$\beta$ emission as likely Lyman continuum leakers}. These galaxies are good candidates for followup studies with current and future telescopes for further observation to determine the presence of Ly$\alpha$ emission and to provide more insight into the properties of the sources that may have contributed to reionisation.
  \item We further investigate the properties of these possible Lyman continuum leakers with boosted nebular emission and high SFR densities. These galaxies and others like them present an opportunity for followup with \textit{JWST}, Gemini, Keck, and the Atacama Large Millimeter/submillimeter Array (ALMA): they are compact, highly star-forming, have larger apparent brightness due to lensing (and are thus easier for followup), and are likely important for reionisation if found to be Lyman continuum leakers. As part of a sub-$L^*$, medium-bright sample, further investigations of these galaxies will also complement previous studies on the determination of the objects responsible for reionising the universe.
\end{itemize}

Further exploration of these and other high-redshift galaxies, especially with space-based Ly$\alpha$ observations with \textit{JWST}, where skylines are not an issue, and rest-frame optical observations, will yield tighter constraints on measurements of their properties. Measurements of Ly$\alpha$ strength can lend insight into Ly$\alpha$ escape, which is correlated with Lyman continuum escape, and \textit{JWST} can also provide higher resolution imaging to yield better size measurements. In addition, measurements of nebular emission lines would provide a better understanding of the ionisation radiation fields strengths of [OIII] and [OII], and H$\alpha$ measurements would provide ionising photon production efficiencies. The detailed information from these observations could lend significant insight into the Epoch of Reionisation and the role of objects in reioinsing the early universe.

\section*{Acknowledgements}

The authors are gratefully acknowledging  Takatoshi Shibuya for providing us the size measurements for z$\sim$7 field galaxies. This material is based upon work supported by NASA/\textit{HST} grant HST-GO-14096, HST-GO-15920, NASA through grant NNX14AN73H , through an award issued by JPL/Caltech, and the National Science Foundation under Grant No. AST-1815458. Based on observations made with the NASA/ESA Hubble Space Telescope, obtained at the Space Telescope Science Institute, which is operated by the Association of Universities for Research in Astronomy, Inc., under NASA contract NAS5-26555. Also based on observations made with the Spitzer Space Telescope, which is operated by the Jet Propulsion Laboratory, California Institute of Technology under a contract with NASA. MB also acknowledges support by the Slovenian national research agency ARRS through grant N1-0238.

\section*{Data Availability}

 The data presented in this paper were obtained from the Multimission Archive at the Space Telescope Science Institution (MAST). StScI is operated by the Association of Universities for Research in Astronomy, Inc., under NASA construct NAS5-26555. Support for MAST and non-HST data is provided by the NASA Office of Space Science via grant NAG5-7584 and other grants and contracts.



\bibliographystyle{mnras}
\bibliography{neufeld} 

\begin{thebibliography}{}
\makeatletter
\relax
\def\mn@urlcharsother{\let\do\@makeother \do\$\do\&\do\#\do\^\do\_\do\%\do\~}
\def\mn@doi{\begingroup\mn@urlcharsother \@ifnextchar [ {\mn@doi@}
  {\mn@doi@[]}}
\def\mn@doi@[#1]#2{\def\@tempa{#1}\ifx\@tempa\@empty \href
  {http://dx.doi.org/#2} {doi:#2}\else \href {http://dx.doi.org/#2} {#1}\fi
  \endgroup}
\def\mn@eprint#1#2{\mn@eprint@#1:#2::\@nil}
\def\mn@eprint@arXiv#1{\href {http://arxiv.org/abs/#1} {{\tt arXiv:#1}}}
\def\mn@eprint@dblp#1{\href {http://dblp.uni-trier.de/rec/bibtex/#1.xml}
  {dblp:#1}}
\def\mn@eprint@#1:#2:#3:#4\@nil{\def\@tempa {#1}\def\@tempb {#2}\def\@tempc
  {#3}\ifx \@tempc \@empty \let \@tempc \@tempb \let \@tempb \@tempa \fi \ifx
  \@tempb \@empty \def\@tempb {arXiv}\fi \@ifundefined
  {mn@eprint@\@tempb}{\@tempb:\@tempc}{\expandafter \expandafter \csname
  mn@eprint@\@tempb\endcsname \expandafter{\@tempc}}}

\bibitem[\protect\citeauthoryear{{Acebron} et~al.,}{{Acebron}
  et~al.}{2018}]{Acebron2018}
{Acebron} A.,  et~al., 2018, \mn@doi [\apj] {10.3847/1538-4357/aabe29}, \href
  {https://ui.adsabs.harvard.edu/abs/2018ApJ...858...42A} {858, 42}

\bibitem[\protect\citeauthoryear{{Acebron} et~al.,}{{Acebron}
  et~al.}{2019a}]{Acebron2019b}
{Acebron} A.,  et~al., 2019a, arXiv e-prints, \href
  {https://ui.adsabs.harvard.edu/abs/2019arXiv191202702A} {p. arXiv:1912.02702}

\bibitem[\protect\citeauthoryear{{Acebron} et~al.,}{{Acebron}
  et~al.}{2019b}]{Acebron2019a}
{Acebron} A.,  et~al., 2019b, \mn@doi [\apj] {10.3847/1538-4357/ab0adf}, \href
  {https://ui.adsabs.harvard.edu/abs/2019ApJ...874..132A} {874, 132}

\bibitem[\protect\citeauthoryear{{Atek} et~al.,}{{Atek}
  et~al.}{2015}]{Atek2015}
{Atek} H.,  et~al., 2015, \mn@doi [\apj] {10.1088/0004-637X/814/1/69}, \href
  {https://ui.adsabs.harvard.edu/abs/2015ApJ...814...69A} {814, 69}

\bibitem[\protect\citeauthoryear{{Atek}, {Richard}, {Kneib}  \&
  {Schaerer}}{{Atek} et~al.}{2018}]{Atek2018}
{Atek} H.,  {Richard} J.,  {Kneib} J.-P.,   {Schaerer} D.,  2018, \mn@doi
  [\mnras] {10.1093/mnras/sty1820}, \href
  {https://ui.adsabs.harvard.edu/abs/2018MNRAS.479.5184A} {479, 5184}

\bibitem[\protect\citeauthoryear{{Bhatawdekar}, {Conselice},
  {Margalef-Bentabol}  \& {Duncan}}{{Bhatawdekar} et~al.}{2019}]{Bhatawdek2019}
{Bhatawdekar} R.,  {Conselice} C.~J.,  {Margalef-Bentabol} B.,   {Duncan} K.,
  2019, \mn@doi [\mnras] {10.1093/mnras/stz866}, \href
  {https://ui.adsabs.harvard.edu/abs/2019MNRAS.486.3805B} {486, 3805}

\bibitem[\protect\citeauthoryear{{Birrer} \& {Amara}}{{Birrer} \&
  {Amara}}{2018}]{Birrer2018}
{Birrer} S.,  {Amara} A.,  2018, {Lenstronomy: Multi-purpose gravitational lens
  modeling software package} (\mn@eprint {ascl} {1804.012})

\bibitem[\protect\citeauthoryear{{Birrer}, {Amara}  \& {Refregier}}{{Birrer}
  et~al.}{2015}]{Birrer2015}
{Birrer} S.,  {Amara} A.,   {Refregier} A.,  2015, \mn@doi [\apj]
  {10.1088/0004-637X/813/2/102}, \href
  {https://ui.adsabs.harvard.edu/abs/2015ApJ...813..102B} {813, 102}

\bibitem[\protect\citeauthoryear{{Bouwens}, {Illingworth}, {Oesch}, {Atek},
  {Lam}  \& {Stefanon}}{{Bouwens} et~al.}{2017}]{Bouwens2017}
{Bouwens} R.~J.,  {Illingworth} G.~D.,  {Oesch} P.~A.,  {Atek} H.,  {Lam} D.,
  {Stefanon} M.,  2017, \mn@doi [\apj] {10.3847/1538-4357/aa74e4}, \href
  {https://ui.adsabs.harvard.edu/abs/2017ApJ...843...41B} {843, 41}

\bibitem[\protect\citeauthoryear{{Bouwens}, {Illingworth}, {van Dokkum},
  {Ribeiro}, {Oesch}  \& {Stefanon}}{{Bouwens} et~al.}{2021}]{Bouwens2021b}
{Bouwens} R.~J.,  {Illingworth} G.~D.,  {van Dokkum} P.~G.,  {Ribeiro} B.,
  {Oesch} P.~A.,   {Stefanon} M.,  2021, \mn@doi [\aj]
  {10.3847/1538-3881/abfda6}, \href
  {https://ui.adsabs.harvard.edu/abs/2021AJ....162..255B} {162, 255}

\bibitem[\protect\citeauthoryear{{Bouwens}, {Illingworth}, {van Dokkum},
  {Oesch}, {Stefanon}  \& {Ribeiro}}{{Bouwens} et~al.}{2022}]{Bouwens2021}
{Bouwens} R.~J.,  {Illingworth} G.~D.,  {van Dokkum} P.~G.,  {Oesch} P.~A.,
  {Stefanon} M.,   {Ribeiro} B.,  2022, \mn@doi [\apj]
  {10.3847/1538-4357/ac4791}, \href
  {https://ui.adsabs.harvard.edu/abs/2022ApJ...927...81B} {927, 81}

\bibitem[\protect\citeauthoryear{{Brammer}, {van Dokkum}  \& {Coppi}}{{Brammer}
  et~al.}{2008}]{Brammer2008}
{Brammer} G.~B.,  {van Dokkum} P.~G.,   {Coppi} P.,  2008, \mn@doi [\apj]
  {10.1086/591786}, \href
  {https://ui.adsabs.harvard.edu/abs/2008ApJ...686.1503B} {686, 1503}

\bibitem[\protect\citeauthoryear{{Buitrago}, {Trujillo}, {Conselice}  \&
  {H{\"a}u{\ss}ler}}{{Buitrago} et~al.}{2013}]{Buitrago2013}
{Buitrago} F.,  {Trujillo} I.,  {Conselice} C.~J.,   {H{\"a}u{\ss}ler} B.,
  2013, \mn@doi [\mnras] {10.1093/mnras/sts124}, \href
  {https://ui.adsabs.harvard.edu/abs/2013MNRAS.428.1460B} {428, 1460}

\bibitem[\protect\citeauthoryear{{Cardamone} et~al.,}{{Cardamone}
  et~al.}{2009}]{cardamone2009}
{Cardamone} C.,  et~al., 2009, \mn@doi [\mnras]
  {10.1111/j.1365-2966.2009.15383.x}, \href
  {https://ui.adsabs.harvard.edu/abs/2009MNRAS.399.1191C} {399, 1191}

\bibitem[\protect\citeauthoryear{{Carrasco}, {Trenti}, {Mutch}  \&
  {Oesch}}{{Carrasco} et~al.}{2018}]{Carrasco2018}
{Carrasco} D.,  {Trenti} M.,  {Mutch} S.,   {Oesch} P.~A.,  2018, \mn@doi
  [\pasa] {10.1017/pasa.2018.17}, \href
  {https://ui.adsabs.harvard.edu/abs/2018PASA...35...22C} {35, e022}

\bibitem[\protect\citeauthoryear{{Cerny} et~al.,}{{Cerny}
  et~al.}{2018}]{Cerny2018}
{Cerny} C.,  et~al., 2018, \mn@doi [\apj] {10.3847/1538-4357/aabe7b}, \href
  {https://ui.adsabs.harvard.edu/abs/2018ApJ...859..159C} {859, 159}

\bibitem[\protect\citeauthoryear{{Chisholm}, {Orlitov{\'a}}, {Schaerer},
  {Verhamme}, {Worseck}, {Izotov}, {Thuan}  \& {Guseva}}{{Chisholm}
  et~al.}{2017}]{Chisholm2017}
{Chisholm} J.,  {Orlitov{\'a}} I.,  {Schaerer} D.,  {Verhamme} A.,  {Worseck}
  G.,  {Izotov} Y.~I.,  {Thuan} T.~X.,   {Guseva} N.~G.,  2017, \mn@doi [\aap]
  {10.1051/0004-6361/201730610}, \href
  {https://ui.adsabs.harvard.edu/abs/2017A&A...605A..67C} {605, A67}

\bibitem[\protect\citeauthoryear{{Cibirka} et~al.,}{{Cibirka}
  et~al.}{2018}]{Cibirka2018}
{Cibirka} N.,  et~al., 2018, \mn@doi [\apj] {10.3847/1538-4357/aad2d3}, \href
  {https://ui.adsabs.harvard.edu/abs/2018ApJ...863..145C} {863, 145}

\bibitem[\protect\citeauthoryear{{Cicone}, {Maiolino}  \& {Marconi}}{{Cicone}
  et~al.}{2016}]{Cicone2016}
{Cicone} C.,  {Maiolino} R.,   {Marconi} A.,  2016, \mn@doi [\aap]
  {10.1051/0004-6361/201424514}, \href
  {https://ui.adsabs.harvard.edu/abs/2016A&A...588A..41C} {588, A41}

\bibitem[\protect\citeauthoryear{{Coe} et~al.,}{{Coe} et~al.}{2013}]{Coe2013}
{Coe} D.,  et~al., 2013, \mn@doi [\apj] {10.1088/0004-637X/762/1/32}, \href
  {https://ui.adsabs.harvard.edu/abs/2013ApJ...762...32C} {762, 32}

\bibitem[\protect\citeauthoryear{{Coe} et~al.,}{{Coe} et~al.}{2019}]{Coe2019}
{Coe} D.,  et~al., 2019, \mn@doi [\apj] {10.3847/1538-4357/ab412b}, \href
  {https://ui.adsabs.harvard.edu/abs/2019ApJ...884...85C} {884, 85}

\bibitem[\protect\citeauthoryear{Dayal}{Dayal}{2019}]{dayal2019}
Dayal P.,  2019, \mn@doi [Proceedings of the International Astronomical Union]
  {10.1017/s1743921320001106}, 15, 43

\bibitem[\protect\citeauthoryear{{Endsley}, {Stark}, {Chevallard}  \&
  {Charlot}}{{Endsley} et~al.}{2021}]{Endsley2021}
{Endsley} R.,  {Stark} D.~P.,  {Chevallard} J.,   {Charlot} S.,  2021, \mn@doi
  [\mnras] {10.1093/mnras/staa3370}, \href
  {https://ui.adsabs.harvard.edu/abs/2021MNRAS.500.5229E} {500, 5229}

\bibitem[\protect\citeauthoryear{{Finkelstein} et~al.,}{{Finkelstein}
  et~al.}{2019}]{Finkelstein2019}
{Finkelstein} S.~L.,  et~al., 2019, \mn@doi [\apj] {10.3847/1538-4357/ab1ea8},
  \href {https://ui.adsabs.harvard.edu/abs/2019ApJ...879...36F} {879, 36}

\bibitem[\protect\citeauthoryear{{Fletcher}, {Tang}, {Robertson}, {Nakajima},
  {Ellis}, {Stark}  \& {Inoue}}{{Fletcher} et~al.}{2019}]{Fletcher2019}
{Fletcher} T.~J.,  {Tang} M.,  {Robertson} B.~E.,  {Nakajima} K.,  {Ellis}
  R.~S.,  {Stark} D.~P.,   {Inoue} A.,  2019, \mn@doi [\apj]
  {10.3847/1538-4357/ab2045}, \href
  {https://ui.adsabs.harvard.edu/abs/2019ApJ...878...87F} {878, 87}

\bibitem[\protect\citeauthoryear{{Flury} et~al.,}{{Flury}
  et~al.}{2022}]{Flury2022}
{Flury} S.~R.,  et~al., 2022, arXiv e-prints, \href
  {https://ui.adsabs.harvard.edu/abs/2022arXiv220111716F} {p. arXiv:2201.11716}

\bibitem[\protect\citeauthoryear{{Franx}, {van Dokkum}, {F{\"o}rster
  Schreiber}, {Wuyts}, {Labb{\'e}}  \& {Toft}}{{Franx}
  et~al.}{2008}]{Franx2008}
{Franx} M.,  {van Dokkum} P.~G.,  {F{\"o}rster Schreiber} N.~M.,  {Wuyts} S.,
  {Labb{\'e}} I.,   {Toft} S.,  2008, \mn@doi [\apj] {10.1086/592431}, \href
  {https://ui.adsabs.harvard.edu/abs/2008ApJ...688..770F} {688, 770}

\bibitem[\protect\citeauthoryear{{Fuller} et~al.,}{{Fuller}
  et~al.}{2020}]{Fuller2020}
{Fuller} S.,  et~al., 2020, \mn@doi [\apj] {10.3847/1538-4357/ab959f}, \href
  {https://ui.adsabs.harvard.edu/abs/2020ApJ...896..156F} {896, 156}

\bibitem[\protect\citeauthoryear{{Grazian} et~al.,}{{Grazian}
  et~al.}{2011}]{Grazian2011}
{Grazian} A.,  et~al., 2011, \mn@doi [\aap] {10.1051/0004-6361/201015754},
  \href {https://ui.adsabs.harvard.edu/abs/2011A&A...532A..33G} {532, A33}

\bibitem[\protect\citeauthoryear{{Grazian} et~al.,}{{Grazian}
  et~al.}{2012}]{Grazian2012}
{Grazian} A.,  et~al., 2012, \mn@doi [\aap] {10.1051/0004-6361/201219669},
  \href {https://ui.adsabs.harvard.edu/abs/2012A&A...547A..51G} {547, A51}

\bibitem[\protect\citeauthoryear{{Heckman}, {Alexandroff}, {Borthakur},
  {Overzier}  \& {Leitherer}}{{Heckman} et~al.}{2015}]{Heckman2015}
{Heckman} T.~M.,  {Alexandroff} R.~M.,  {Borthakur} S.,  {Overzier} R.,
  {Leitherer} C.,  2015, \mn@doi [\apj] {10.1088/0004-637X/809/2/147}, \href
  {https://ui.adsabs.harvard.edu/abs/2015ApJ...809..147H} {809, 147}

\bibitem[\protect\citeauthoryear{{Hoag} et~al.,}{{Hoag}
  et~al.}{2019}]{hoag2019b}
{Hoag} A.,  et~al., 2019, \mn@doi [\mnras] {10.1093/mnras/stz1768}, \href
  {https://ui.adsabs.harvard.edu/abs/2019MNRAS.488..706H} {488, 706}

\bibitem[\protect\citeauthoryear{{Holwerda}, {Bouwens}, {Oesch}, {Smit},
  {Illingworth}  \& {Labbe}}{{Holwerda} et~al.}{2015}]{Holwerda2015}
{Holwerda} B.~W.,  {Bouwens} R.,  {Oesch} P.,  {Smit} R.,  {Illingworth} G.,
  {Labbe} I.,  2015, \mn@doi [\apj] {10.1088/0004-637X/808/1/6}, \href
  {https://ui.adsabs.harvard.edu/abs/2015ApJ...808....6H} {808, 6}

\bibitem[\protect\citeauthoryear{{Huang}, {Ferguson}, {Ravindranath}  \&
  {Su}}{{Huang} et~al.}{2013}]{Huang2013}
{Huang} K.-H.,  {Ferguson} H.~C.,  {Ravindranath} S.,   {Su} J.,  2013, \mn@doi
  [\apj] {10.1088/0004-637X/765/1/68}, \href
  {https://ui.adsabs.harvard.edu/abs/2013ApJ...765...68H} {765, 68}

\bibitem[\protect\citeauthoryear{{Huang} et~al.,}{{Huang}
  et~al.}{2017}]{Huang2017}
{Huang} K.-H.,  et~al., 2017, \mn@doi [\apj] {10.3847/1538-4357/aa62a6}, \href
  {https://ui.adsabs.harvard.edu/abs/2017ApJ...838....6H} {838, 6}

\bibitem[\protect\citeauthoryear{{Ishigaki}, {Kawamata}, {Ouchi}, {Oguri},
  {Shimasaku}  \& {Ono}}{{Ishigaki} et~al.}{2018}]{Ishigaki2018}
{Ishigaki} M.,  {Kawamata} R.,  {Ouchi} M.,  {Oguri} M.,  {Shimasaku} K.,
  {Ono} Y.,  2018, \mn@doi [\apj] {10.3847/1538-4357/aaa544}, \href
  {https://ui.adsabs.harvard.edu/abs/2018ApJ...854...73I} {854, 73}

\bibitem[\protect\citeauthoryear{Izotov, Schaerer, Thuan, Worseck, Guseva,
  Orlitová  \& Verhamme}{Izotov et~al.}{2016}]{Izotov2016b}
Izotov Y.~I.,  Schaerer D.,  Thuan T.~X.,  Worseck G.,  Guseva N.~G.,
  Orlitová I.,   Verhamme A.,  2016, \mn@doi [\mnras] {10.1093/mnras/stw1205},
  461, 3683

\bibitem[\protect\citeauthoryear{{Izotov}, {Worseck}, {Schaerer}, {Guseva},
  {Chisholm}, {Thuan}, {Fricke}  \& {Verhamme}}{{Izotov}
  et~al.}{2021}]{Izotov2021}
{Izotov} Y.~I.,  {Worseck} G.,  {Schaerer} D.,  {Guseva} N.~G.,  {Chisholm} J.,
   {Thuan} T.~X.,  {Fricke} K.~J.,   {Verhamme} A.,  2021, \mn@doi [\mnras]
  {10.1093/mnras/stab612}, \href
  {https://ui.adsabs.harvard.edu/abs/2021MNRAS.503.1734I} {503, 1734}

\bibitem[\protect\citeauthoryear{{Jullo} \& {Kneib}}{{Jullo} \&
  {Kneib}}{2009}]{Jullo2009}
{Jullo} E.,  {Kneib} J.-P.,  2009, \mn@doi [\mnras]
  {10.1111/j.1365-2966.2009.14654.x}, \href
  {http://adsabs.harvard.edu/abs/2009MNRAS.395.1319J} {395, 1319}

\bibitem[\protect\citeauthoryear{{Kawamata}, {Ishigaki}, {Shimasaku}, {Oguri}
  \& {Ouchi}}{{Kawamata} et~al.}{2015}]{Kawamata2015}
{Kawamata} R.,  {Ishigaki} M.,  {Shimasaku} K.,  {Oguri} M.,   {Ouchi} M.,
  2015, \mn@doi [\apj] {10.1088/0004-637X/804/2/103}, \href
  {https://ui.adsabs.harvard.edu/abs/2015ApJ...804..103K} {804, 103}

\bibitem[\protect\citeauthoryear{{Kawamata}, {Ishigaki}, {Shimasaku}, {Oguri},
  {Ouchi}  \& {Tanigawa}}{{Kawamata} et~al.}{2018}]{Kawamata2018}
{Kawamata} R.,  {Ishigaki} M.,  {Shimasaku} K.,  {Oguri} M.,  {Ouchi} M.,
  {Tanigawa} S.,  2018, \mn@doi [\apj] {10.3847/1538-4357/aaa6cf}, \href
  {https://ui.adsabs.harvard.edu/abs/2018ApJ...855....4K} {855, 4}

\bibitem[\protect\citeauthoryear{{Kennicutt}}{{Kennicutt}}{1998}]{Kennicutt1998}
{Kennicutt} Robert~C. J.,  1998, \mn@doi [\araa]
  {10.1146/annurev.astro.36.1.189}, \href
  {https://ui.adsabs.harvard.edu/abs/1998ARA&A..36..189K} {36, 189}

\bibitem[\protect\citeauthoryear{{Kim}, {Malhotra}, {Rhoads}  \& {Yang}}{{Kim}
  et~al.}{2021}]{Kim2021}
{Kim} K.~J.,  {Malhotra} S.,  {Rhoads} J.~E.,   {Yang} H.,  2021, \mn@doi
  [\apj] {10.3847/1538-4357/abf833}, \href
  {https://ui.adsabs.harvard.edu/abs/2021ApJ...914....2K} {914, 2}

\bibitem[\protect\citeauthoryear{{Krywult} et~al.,}{{Krywult}
  et~al.}{2017}]{Krywult2017}
{Krywult} J.,  et~al., 2017, \mn@doi [\aap] {10.1051/0004-6361/201628953},
  \href {https://ui.adsabs.harvard.edu/abs/2017A&A...598A.120K} {598, A120}

\bibitem[\protect\citeauthoryear{{Leethochawalit}, {Trenti}, {Morishita},
  {Roberts-Borsani}  \& {Treu}}{{Leethochawalit} et~al.}{2022}]{Leet2022}
{Leethochawalit} N.,  {Trenti} M.,  {Morishita} T.,  {Roberts-Borsani} G.,
  {Treu} T.,  2022, \mn@doi [\mnras] {10.1093/mnras/stab3265}, \href
  {https://ui.adsabs.harvard.edu/abs/2022MNRAS.509.5836L} {509, 5836}

\bibitem[\protect\citeauthoryear{{Lemaux} et~al.,}{{Lemaux}
  et~al.}{2021}]{Lemaux2021}
{Lemaux} B.~C.,  et~al., 2021, \mn@doi [\mnras] {10.1093/mnras/stab924}, \href
  {https://ui.adsabs.harvard.edu/abs/2021MNRAS.504.3662L} {504, 3662}

\bibitem[\protect\citeauthoryear{{Livermore}, {Finkelstein}  \&
  {Lotz}}{{Livermore} et~al.}{2017}]{Livermore2017}
{Livermore} R.~C.,  {Finkelstein} S.~L.,   {Lotz} J.~M.,  2017, \mn@doi [\apj]
  {10.3847/1538-4357/835/2/113}, \href
  {https://ui.adsabs.harvard.edu/abs/2017ApJ...835..113L} {835, 113}

\bibitem[\protect\citeauthoryear{{Lotz} et~al.,}{{Lotz}
  et~al.}{2017}]{Lotz2017}
{Lotz} J.~M.,  et~al., 2017, \mn@doi [\apj] {10.3847/1538-4357/837/1/97}, \href
  {https://ui.adsabs.harvard.edu/abs/2017ApJ...837...97L} {837, 97}

\bibitem[\protect\citeauthoryear{{MacKenty}, {Kimble}, {O'Connell}  \&
  {Townsend}}{{MacKenty} et~al.}{2008}]{Mackenty2008}
{MacKenty} J.~W.,  {Kimble} R.~A.,  {O'Connell} R.~W.,   {Townsend} J.~A.,
  2008, in {Oschmann} Jacobus~M. J.,  {de Graauw} M. W.~M.,   {MacEwen} H.~A.,
  eds,  Society of Photo-Optical Instrumentation Engineers (SPIE) Conference
  Series Vol. 7010, Space Telescopes and Instrumentation 2008: Optical,
  Infrared, and Millimeter. p. 70101F, \mn@doi{10.1117/12.790039}

\bibitem[\protect\citeauthoryear{{Malkan} \& {Malkan}}{{Malkan} \&
  {Malkan}}{2021}]{Malkan2021}
{Malkan} M.~A.,  {Malkan} B.~K.,  2021, \mn@doi [\apj]
  {10.3847/1538-4357/abd84e}, \href
  {https://ui.adsabs.harvard.edu/abs/2021ApJ...909...92M} {909, 92}

\bibitem[\protect\citeauthoryear{{Marchi} et~al.,}{{Marchi}
  et~al.}{2018}]{Marchi2018}
{Marchi} F.,  et~al., 2018, \mn@doi [\aap] {10.1051/0004-6361/201732133}, \href
  {https://ui.adsabs.harvard.edu/abs/2018A&A...614A..11M} {614, A11}

\bibitem[\protect\citeauthoryear{{Matthee} et~al.,}{{Matthee}
  et~al.}{2019}]{Matthee2019}
{Matthee} J.,  et~al., 2019, \mn@doi [\apj] {10.3847/1538-4357/ab2f81}, \href
  {https://ui.adsabs.harvard.edu/abs/2019ApJ...881..124M} {881, 124}

\bibitem[\protect\citeauthoryear{{Matthee} et~al.,}{{Matthee}
  et~al.}{2021}]{Mathee2021}
{Matthee} J.,  et~al., 2021, arXiv e-prints, \href
  {https://ui.adsabs.harvard.edu/abs/2021arXiv211011967M} {p. arXiv:2110.11967}

\bibitem[\protect\citeauthoryear{{Morishita}, {Ichikawa}  \&
  {Kajisawa}}{{Morishita} et~al.}{2014}]{Morishita2014}
{Morishita} T.,  {Ichikawa} T.,   {Kajisawa} M.,  2014, \mn@doi [\apj]
  {10.1088/0004-637X/785/1/18}, \href
  {https://ui.adsabs.harvard.edu/abs/2014ApJ...785...18M} {785, 18}

\bibitem[\protect\citeauthoryear{{Mowla} et~al.,}{{Mowla}
  et~al.}{2019}]{Mowla2019}
{Mowla} L.~A.,  et~al., 2019, \mn@doi [\apj] {10.3847/1538-4357/ab290a}, \href
  {https://ui.adsabs.harvard.edu/abs/2019ApJ...880...57M} {880, 57}

\bibitem[\protect\citeauthoryear{{Naidu}, {Tacchella}, {Mason}, {Bose}, {Oesch}
   \& {Conroy}}{{Naidu} et~al.}{2020}]{Naidu2020}
{Naidu} R.~P.,  {Tacchella} S.,  {Mason} C.~A.,  {Bose} S.,  {Oesch} P.~A.,
  {Conroy} C.,  2020, \mn@doi [\apj] {10.3847/1538-4357/ab7cc9}, \href
  {https://ui.adsabs.harvard.edu/abs/2020ApJ...892..109N} {892, 109}

\bibitem[\protect\citeauthoryear{{Naidu} et~al.,}{{Naidu}
  et~al.}{2021}]{Naidu2021}
{Naidu} R.~P.,  et~al., 2021, arXiv e-prints, \href
  {https://ui.adsabs.harvard.edu/abs/2021arXiv211011961N} {p. arXiv:2110.11961}

\bibitem[\protect\citeauthoryear{{Oguri}}{{Oguri}}{2010}]{Oguri2010}
{Oguri} M.,  2010, \mn@doi [\pasj] {10.1093/pasj/62.4.1017}, \href
  {http://adsabs.harvard.edu/abs/2010PASJ...62.1017O} {62, 1017}

\bibitem[\protect\citeauthoryear{{Okabe} et~al.,}{{Okabe}
  et~al.}{2020}]{Okabe2020}
{Okabe} T.,  et~al., 2020, \mn@doi [\mnras] {10.1093/mnras/staa1479}, \href
  {https://ui.adsabs.harvard.edu/abs/2020MNRAS.496.2591O} {496, 2591}

\bibitem[\protect\citeauthoryear{{Oke}}{{Oke}}{1974}]{Oke1974}
{Oke} J.~B.,  1974, \mn@doi [\apjs] {10.1086/190287}, \href
  {http://adsabs.harvard.edu/abs/1974ApJS...27...21O} {27, 21}

\bibitem[\protect\citeauthoryear{{Ono} et~al.,}{{Ono} et~al.}{2013}]{Ono2013}
{Ono} Y.,  et~al., 2013, \mn@doi [\apj] {10.1088/0004-637X/777/2/155}, \href
  {https://ui.adsabs.harvard.edu/abs/2013ApJ...777..155O} {777, 155}

\bibitem[\protect\citeauthoryear{{Paterno-Mahler} et~al.,}{{Paterno-Mahler}
  et~al.}{2018}]{PaternoMahler2018}
{Paterno-Mahler} R.,  et~al., 2018, \mn@doi [\apj] {10.3847/1538-4357/aad239},
  \href {http://adsabs.harvard.edu/abs/2018ApJ...863..154P} {863, 154}

\bibitem[\protect\citeauthoryear{{Pelliccia} et~al.,}{{Pelliccia}
  et~al.}{2021}]{Pelliccia2021}
{Pelliccia} D.,  et~al., 2021, \mn@doi [\apjl] {10.3847/2041-8213/abdf56},
  \href {https://ui.adsabs.harvard.edu/abs/2021ApJ...908L..30P} {908, L30}

\bibitem[\protect\citeauthoryear{{Pentericci} et~al.,}{{Pentericci}
  et~al.}{2014}]{pentericci2014}
{Pentericci} L.,  et~al., 2014, \mn@doi [\apj] {10.1088/0004-637X/793/2/113},
  \href {https://ui.adsabs.harvard.edu/abs/2014ApJ...793..113P} {793, 113}

\bibitem[\protect\citeauthoryear{{Petty}, {de Mello}, {Gallagher}, {Gardner},
  {Lotz}, {Mountain}  \& {Smith}}{{Petty} et~al.}{2009}]{Petty2009}
{Petty} S.~M.,  {de Mello} D.~F.,  {Gallagher} John~S. I.,  {Gardner} J.~P.,
  {Lotz} J.~M.,  {Mountain} C.~M.,   {Smith} L.~J.,  2009, \mn@doi [\aj]
  {10.1088/0004-6256/138/2/362}, \href
  {https://ui.adsabs.harvard.edu/abs/2009AJ....138..362P} {138, 362}

\bibitem[\protect\citeauthoryear{{Postman} et~al.,}{{Postman}
  et~al.}{2012}]{Postman2012}
{Postman} M.,  et~al., 2012, \mn@doi [\apjs] {10.1088/0067-0049/199/2/25},
  \href {https://ui.adsabs.harvard.edu/abs/2012ApJS..199...25P} {199, 25}

\bibitem[\protect\citeauthoryear{{Ribeiro} et~al.,}{{Ribeiro}
  et~al.}{2016}]{Ribeiro2016}
{Ribeiro} B.,  et~al., 2016, \mn@doi [\aap] {10.1051/0004-6361/201628249},
  \href {https://ui.adsabs.harvard.edu/abs/2016A&A...593A..22R} {593, A22}

\bibitem[\protect\citeauthoryear{{Roberts-Borsani} et~al.,}{{Roberts-Borsani}
  et~al.}{2016}]{borsani2016}
{Roberts-Borsani} G.~W.,  et~al., 2016, \mn@doi [\apj]
  {10.3847/0004-637X/823/2/143}, \href
  {https://ui.adsabs.harvard.edu/abs/2016ApJ...823..143R} {823, 143}

\bibitem[\protect\citeauthoryear{{Roberts-Borsani}, {Ellis}  \&
  {Laporte}}{{Roberts-Borsani} et~al.}{2020}]{borsani2020}
{Roberts-Borsani} G.~W.,  {Ellis} R.~S.,   {Laporte} N.,  2020, \mn@doi
  [\mnras] {10.1093/mnras/staa2085}, \href
  {https://ui.adsabs.harvard.edu/abs/2020MNRAS.497.3440R} {497, 3440}

\bibitem[\protect\citeauthoryear{{Robertson}}{{Robertson}}{2021}]{robertson2021}
{Robertson} B.~E.,  2021, arXiv e-prints, \href
  {https://ui.adsabs.harvard.edu/abs/2021arXiv211013160R} {p. arXiv:2110.13160}

\bibitem[\protect\citeauthoryear{{Salmon} et~al.,}{{Salmon}
  et~al.}{2020}]{Salmon2020}
{Salmon} B.,  et~al., 2020, \mn@doi [\apj] {10.3847/1538-4357/ab5a8b}, \href
  {https://ui.adsabs.harvard.edu/abs/2020ApJ...889..189S} {889, 189}

\bibitem[\protect\citeauthoryear{{Saxena} et~al.,}{{Saxena}
  et~al.}{2021}]{Saxena2021}
{Saxena} A.,  et~al., 2021, arXiv e-prints, \href
  {https://ui.adsabs.harvard.edu/abs/2021arXiv210903662S} {p. arXiv:2109.03662}

\bibitem[\protect\citeauthoryear{{Shapley}, {Steidel}, {Strom},
  {Bogosavljevi{\'c}}, {Reddy}, {Siana}, {Mostardi}  \& {Rudie}}{{Shapley}
  et~al.}{2016}]{Shapley2016}
{Shapley} A.~E.,  {Steidel} C.~C.,  {Strom} A.~L.,  {Bogosavljevi{\'c}} M.,
  {Reddy} N.~A.,  {Siana} B.,  {Mostardi} R.~E.,   {Rudie} G.~C.,  2016,
  \mn@doi [\apjl] {10.3847/2041-8205/826/2/L24}, \href
  {https://ui.adsabs.harvard.edu/abs/2016ApJ...826L..24S} {826, L24}

\bibitem[\protect\citeauthoryear{{Shen}, {Hopkins}, {Faucher-Gigu{\`e}re},
  {Alexander}, {Richards}, {Ross}  \& {Hickox}}{{Shen} et~al.}{2020}]{Shen2020}
{Shen} X.,  {Hopkins} P.~F.,  {Faucher-Gigu{\`e}re} C.-A.,  {Alexander} D.~M.,
  {Richards} G.~T.,  {Ross} N.~P.,   {Hickox} R.~C.,  2020, \mn@doi [\mnras]
  {10.1093/mnras/staa1381}, \href
  {https://ui.adsabs.harvard.edu/abs/2020MNRAS.495.3252S} {495, 3252}

\bibitem[\protect\citeauthoryear{{Shibuya}, {Ouchi}  \& {Harikane}}{{Shibuya}
  et~al.}{2015}]{Shibuya2015}
{Shibuya} T.,  {Ouchi} M.,   {Harikane} Y.,  2015, \mn@doi [\apjs]
  {10.1088/0067-0049/219/2/15}, \href
  {https://ui.adsabs.harvard.edu/abs/2015ApJS..219...15S} {219, 15}

\bibitem[\protect\citeauthoryear{{Shibuya}, {Ouchi}, {Harikane}  \&
  {Nakajima}}{{Shibuya} et~al.}{2019}]{Shibuya2019}
{Shibuya} T.,  {Ouchi} M.,  {Harikane} Y.,   {Nakajima} K.,  2019, \mn@doi
  [\apj] {10.3847/1538-4357/aaf64b}, \href
  {https://ui.adsabs.harvard.edu/abs/2019ApJ...871..164S} {871, 164}

\bibitem[\protect\citeauthoryear{{Smit} et~al.,}{{Smit}
  et~al.}{2015}]{Smit2015}
{Smit} R.,  et~al., 2015, \mn@doi [\apj] {10.1088/0004-637X/801/2/122}, \href
  {https://ui.adsabs.harvard.edu/abs/2015ApJ...801..122S} {801, 122}

\bibitem[\protect\citeauthoryear{{Strait} et~al.,}{{Strait}
  et~al.}{2020}]{Strait2020}
{Strait} V.,  et~al., 2020, \mn@doi [\apj] {10.3847/1538-4357/ab5daf}, \href
  {https://ui.adsabs.harvard.edu/abs/2020ApJ...888..124S} {888, 124}

\bibitem[\protect\citeauthoryear{{Strait} et~al.,}{{Strait}
  et~al.}{2021}]{Strait2021}
{Strait} V.,  et~al., 2021, \mn@doi [\apj] {10.3847/1538-4357/abe533}, \href
  {https://ui.adsabs.harvard.edu/abs/2021ApJ...910..135S} {910, 135}

\bibitem[\protect\citeauthoryear{{Tang}, {Stark}, {Chevallard}  \&
  {Charlot}}{{Tang} et~al.}{2019}]{tang2019}
{Tang} M.,  {Stark} D.~P.,  {Chevallard} J.,   {Charlot} S.,  2019, \mn@doi
  [\mnras] {10.1093/mnras/stz2236}, \href
  {https://ui.adsabs.harvard.edu/abs/2019MNRAS.489.2572T} {489, 2572}

\bibitem[\protect\citeauthoryear{{Trenti} \& {Stiavelli}}{{Trenti} \&
  {Stiavelli}}{2008}]{Trenti2008}
{Trenti} M.,  {Stiavelli} M.,  2008, \mn@doi [\apj] {10.1086/528674}, \href
  {https://ui.adsabs.harvard.edu/abs/2008ApJ...676..767T} {676, 767}

\bibitem[\protect\citeauthoryear{{Trenti} et~al.,}{{Trenti}
  et~al.}{2011}]{Trenti2011}
{Trenti} M.,  et~al., 2011, \mn@doi [\apjl] {10.1088/2041-8205/727/2/L39},
  \href {https://ui.adsabs.harvard.edu/abs/2011ApJ...727L..39T} {727, L39}

\bibitem[\protect\citeauthoryear{{Vanzella} et~al.,}{{Vanzella}
  et~al.}{2020}]{Vanzella2020}
{Vanzella} E.,  et~al., 2020, \mn@doi [\mnras] {10.1093/mnrasl/slaa041}, \href
  {https://ui.adsabs.harvard.edu/abs/2020MNRAS.494L..81V} {494, L81}

\bibitem[\protect\citeauthoryear{{Verhamme}, {Orlitov{\'a}}, {Schaerer}  \&
  {Hayes}}{{Verhamme} et~al.}{2015}]{Verhamme2015}
{Verhamme} A.,  {Orlitov{\'a}} I.,  {Schaerer} D.,   {Hayes} M.,  2015, \mn@doi
  [\aap] {10.1051/0004-6361/201423978}, \href
  {https://ui.adsabs.harvard.edu/abs/2015A&A...578A...7V} {578, A7}

\bibitem[\protect\citeauthoryear{{Verhamme}, {Orlitov{\'a}}, {Schaerer},
  {Izotov}, {Worseck}, {Thuan}  \& {Guseva}}{{Verhamme}
  et~al.}{2017}]{Verhamme2017}
{Verhamme} A.,  {Orlitov{\'a}} I.,  {Schaerer} D.,  {Izotov} Y.,  {Worseck} G.,
   {Thuan} T.~X.,   {Guseva} N.,  2017, \mn@doi [\aap]
  {10.1051/0004-6361/201629264}, \href
  {https://ui.adsabs.harvard.edu/abs/2017A&A...597A..13V} {597, A13}

\bibitem[\protect\citeauthoryear{{Yang}, {Birrer}  \& {Treu}}{{Yang}
  et~al.}{2020a}]{Yang2020}
{Yang} L.,  {Birrer} S.,   {Treu} T.,  2020a, \mn@doi [\mnras]
  {10.1093/mnras/staa1649}, \href
  {https://ui.adsabs.harvard.edu/abs/2020MNRAS.496.2648Y} {496, 2648}

\bibitem[\protect\citeauthoryear{Yang, Roberts-Borsani, Treu, Birrer, Morishita
   \& Bradač}{Yang et~al.}{2020b}]{Yang2020a}
Yang L.,  Roberts-Borsani G.,  Treu T.,  Birrer S.,  Morishita T.,   Bradač
  M.,  2020b, \mn@doi [\mnras] {10.1093/mnras/staa3713}, 501, 1028

\bibitem[\protect\citeauthoryear{{Yang} et~al.,}{{Yang}
  et~al.}{2022}]{Yang2022}
{Yang} L.,  et~al., 2022, arXiv e-prints, \href
  {https://ui.adsabs.harvard.edu/abs/2022arXiv220108858Y} {p. arXiv:2201.08858}

\bibitem[\protect\citeauthoryear{{Yue} et~al.,}{{Yue} et~al.}{2018}]{Yue2018}
{Yue} B.,  et~al., 2018, \mn@doi [\apj] {10.3847/1538-4357/aae77f}, \href
  {https://ui.adsabs.harvard.edu/abs/2018ApJ...868..115Y} {868, 115}

\bibitem[\protect\citeauthoryear{{Zitrin} et~al.,}{{Zitrin}
  et~al.}{2014}]{Zitrin2014}
{Zitrin} A.,  et~al., 2014, \mn@doi [\apjl] {10.1088/2041-8205/793/1/L12},
  \href {https://ui.adsabs.harvard.edu/abs/2014ApJ...793L..12Z} {793, L12}

\bibitem[\protect\citeauthoryear{{Zitrin} et~al.,}{{Zitrin}
  et~al.}{2017}]{Zitrin2017}
{Zitrin} A.,  et~al., 2017, \mn@doi [\apjl] {10.3847/2041-8213/aa69be}, \href
  {http://adsabs.harvard.edu/abs/2017ApJ...839L..11Z} {839, L11}

\bibitem[\protect\citeauthoryear{{van der Wel} et~al.,}{{van der Wel}
  et~al.}{2014}]{vanderWel2014}
{van der Wel} A.,  et~al., 2014, \mn@doi [\apj] {10.1088/0004-637X/788/1/28},
  \href {https://ui.adsabs.harvard.edu/abs/2014ApJ...788...28V} {788, 28}

\makeatother
\end{thebibliography}





\bsp	
\label{lastpage}
\end{document}